\documentclass[11pt]{article}
\usepackage[utf8]{inputenc}
\usepackage{authblk}
\usepackage{setspace}
\usepackage[margin=1.25in]{geometry}
\usepackage{graphicx}
\graphicspath{ {./figures/} }
\usepackage{subcaption}
\usepackage{amsmath}
\usepackage{hyperref}
\usepackage{xcolor}
\usepackage{latexsym}
\usepackage{amssymb}
\usepackage{amsfonts}
\usepackage{amsthm}
\usepackage{bbm}
\usepackage{color}
\usepackage{cancel} 
\usepackage{ulem}
\usepackage{bm}
\usepackage{soul}
\usepackage{array}
\usepackage{multirow}
\usepackage[all]{xy}
\usepackage[bottom]{footmisc}
\usepackage{svg}
\usepackage{slashed}		

\hypersetup{
	bookmarksnumbered,
	unicode,			
	colorlinks,			
	citecolor=[rgb]{.9,0,.5},	
	urlcolor=[rgb]{0,0,1},	
	linkcolor=[rgb]{0,.7,0}	
	}

\usepackage[
    natbib=true,
    style=numeric-comp,
    sorting=none
]{biblatex}

\catcode`@=11
\def\secteqno{\@addtoreset{equation}{section}%
	\def\theequation{\thesection.\arabic{equation}}}
\catcode`@=12
\secteqno
\def\dd{\hbox{\,\Large$\triangleright$}}
\newcommand{\be}{\begin{equation}}
	\newcommand{\ee}{\end{equation}}
\newcommand{\bea}{\begin{eqnarray}}
	\newcommand{\eea}{\end{eqnarray}}
\newcommand{\bref}[1]{(\ref{#1})}
\newcommand{\nn}{\nonumber}

\def\dig#1{\setbox0=\hbox{$#1M$}
	\hskip.06\wd0 \vrule width.07\wd0 height.63\wd0 depth.01\wd0 
	\vrule width.37\wd0 height.63\wd0 depth-.56\wd0 \hskip-.4\wd0
	\vrule width.25\wd0 height.35\wd0 depth-.28\wd0 
	\vrule width.07\wd0 height.35\wd0 depth-.17\wd0 \hskip.14\wd0}
\def\digamma{{\mathpalette\dig{}}}
\def\={{\;=\;}}\def\+{{\;+\;}}

\def\dig#1{\setbox0=\hbox{$#1M$}
	\hskip.06\wd0 \vrule width.08\wd0 height.63\wd0 depth.01\wd0 
	\vrule width.37\wd0 height.63\wd0 depth-.55\wd0 \hskip-.4\wd0
	\vrule width.25\wd0 height.36\wd0 depth-.28\wd0 
	\vrule width.08\wd0 height.36\wd0 depth-.17\wd0 \hskip.14\wd0}
\def\digamma{{\mathpalette\dig{}}}
\def\bop#1{\setbox0=\hbox{$#1M$}\mkern1.5mu
	\vbox{\hrule height0pt depth.1\ht0
		\hbox{\vrule width.1\ht0 height.8\ht0 \kern.8\ht0
			\vrule width.1\ht0}\hrule height.1\ht0}\mkern1.5mu}

\def\dd{\hbox{\,\Large$\triangleright$}} 

\def\don#1#2{{\buildrel{\mkern2.5mu\raise-.1em\hbox{$\scriptstyle#1$}\mkern-2.5mu}\over{#2}}}	
\def\dron#1#2{{\buildrel{{\raise-.1em\hbox{$\scriptstyle#1$}}}\over{#2}}}		
\newcommand{\ctext}[1]{\raise0.2ex\hbox{\textcircled{\scriptsize{#1}}}}
\newcommand{\Red}[1]{\textcolor{red}{#1}}

\title{Gauss law constraint in ${\cal A}$-theory branes  }

\author[$\heartsuit$$\diamondsuit$]{Machiko Hatsuda\footnote{Email: \href{mhatsuda@juntendo.ac.jp}{mhatsuda@juntendo.ac.jp} }}
\author[$\ddagger$]{Ond\v{r}ej Hul\'{\i}k}
\author[$\#$]{ William D. Linch }
\author[$\clubsuit$]{\\Di Wang}
\author[$\spadesuit$]{ Yu-Ping Wang 
}

\affil[$\heartsuit$]{\textit{Department of Radiological Technology, Faculty of Health Science, Juntendo University
		Yushima, Bunkyou-ku, Tokyo 113-0034, Japan}}
\affil[$\diamondsuit$]{\textit{KEK Theory Center, High Energy Accelerator Research Organization,
		Tsukuba, Ibaraki 305-0801, Japan}}
\affil[$\ddagger$]{\textit{Institute for Mathematics 
 Ruprecht-Karls-Universitat Heidelberg,
 69120 Heidelberg, Germany}}
\affil[$\#$]{\textit{Thomas Jefferson High School for Science and Technology, Alexandria, VA 22312, USA} }
\affil[$\clubsuit$]{\textit{ Department of Physics, SUNY Stony Brook University, Stony Brook, NY 11794, USA}}
\affil[$\spadesuit$]{\textit{String theory group, National Taiwan University, Taipei city, 106 , Taiwan}}
\date{\today}

\begin{document}
	
	\newgeometry{top=0.1in,bottom=1in,right=1in,left=1in}
	\vskip 0.1in
	\hfill KEK-TH-2814
	{\let\newpage\relax\maketitle}
	
\begin{abstract}
${\cal A}$-theory realizes U-duality symmetry by extending the string worldsheet to a higher dimensional brane worldvolume,
in which the worldvolume and the spacetime belong to different representations of the exceptional group.
The closure of the brane Virasoro algebra requires the Gauss law constraint.
The Gauss law constraint promotes spacetime coordinates to gauge fields and extends the string worldsheet into the brane worldvolume. 
While the Virasoro constraint is used to reduce the spacetime coordinate, the Gauss law constraint is used to reduce both  the worldvolume and the spacetime coordinates.
As in conventional gauge theories, the treatment of the Gauss law constraint is a technically important aspect of the quantization of ${\cal A}$-theory.
We show that the string solution is the only consistent solution of the Gauss law dimensional reduction condition for D=3 and 4 cases
 {within the present framework, while leaving open the possibility of higher dimensional branes described by nonlinear actions}.
This result implies that the physical symmetry of the theory is two-dimensional conformal symmetry, suggesting that the theory admits a string-like quantization.
We further construct a string solution that is covariant under the exceptional group symmetry. 
The relation between this solution and the constant charge parameter appearing in the exceptional $\sigma$-model is also discussed.
\end{abstract}
	
\restoregeometry
\tableofcontents
\newpage
\section{Introduction}


\subsection{Overview}

String duality plays a central role in superstring theory. It not only relates the six superstring theories including M-theory, but also provides the foundation for constructing a unified framework. In superstring theories, the gravitational field and gauge fields can be understood as coset parameters of the duality symmetry group, indicating that duality symmetry  determines the structure of spacetime geometry itself.

 The SL(2) S-duality symmetry combines with the spacetime GL(D) symmetry to form the group GL(D+1), while the T-duality of the D-dimensional space is described by the group O(D,D).
These groups, GL(D+1) and O(D,D), are unified into the exceptional group E$_{{\rm D}+1}$ as the U-duality symmetry group.
The exceptional group serves as the guiding symmetry of ${\cal A}$-theory, a brane worldvolume theory with manifest U-duality 
\cite{Polacek:2014cva,Linch:2015fca,Linch:2015lwa,Linch:2015qva,Linch:2016ipx, Linch:2017eru,Siegel:2018puf,Siegel:2019wrr,Siegel:2020qef} (see review \cite{Hatsuda:2021wpb}). 

The  worldvolume of the ${\cal A}$-theory brane is described by the Virasoro constraint, and the closure of the Virasoro algebra requires the Gauss law constraint. The  Gauss law constraint generates the new worldvolume gauge symmetry of the spacetime coordinate.  The gauge parameter $\lambda^{\cal M}$ is also a representation of the exceptional group as well as spacetime coordinate $X^M$ and the worldvolume coordinate $\sigma_m$. The spacetime coordinate $X^M(\sigma_m)$ is a gauge field, and the worldvolume gauge transformation is determined by the Clebsch–Gordan-Wigner (CGW) coefficient that relates the spacetime, worldvolume, and gauge-parameter indices.

As the spacetime is extended to manifest duality symmetry, the additional degrees of freedom are reduced by imposing the selfduality condition.  In six dimensions a covariant selfdual gauge theory can be described by a rank two antisymmetric gauge field 
\cite{Siegel:1983es}, which is used for the D=3 ${\cal A}$-theory five brane \cite{Linch:2015fya}. 
In Hamiltonian formulation, the selfdual field strength on the M5-brane contributes to the SO(5,5) spacetime current  \cite{Hatsuda:2013dya}. 
This structure suggests that the rank two antisymmetric gauge field naturally serves as a U-duality covariant coordinate, thereby encoding the selfduality condition in a covariant manner.
In other words, this implies the existence of a Gauss law constraint that generates the gauge symmetry associated with the extended coordinates.

T-duality of the open string leads to the Dirac-Born-Infeld (DBI) gauge field on D-branes \cite{Polchinski:1996na}.
The DBI electric field, which is the canonical conjugate of the DBI gauge field, is shown to correspond to the fundamental string charge
\cite{Witten:1995im}.
In the Hamiltonian formulation, the Gauss law constraint appears as a secondary constraint and also enters on the right hand side of the supersymmetry algebra
\cite{Hatsuda:1997pq,Kamimura:1997ju,Hatsuda:1998by, Hatsuda:2012uk,Hatsuda:2013dya}.
The local superalgebra shows that the DBI electric field represents the NS–NS charge of D-branes, while the DBI magnetic field is associated with the R–R charge.
This identification implies that the worldvolume derivative encodes the brane charge density.
Consequently, this structure provides a natural basis for understanding the correspondence between the extended worldvolume in our covariantized string solution and the constant charge parameter appearing in the exceptional $\sigma$-model
\cite{Arvanitakis:2017hwb,Arvanitakis:2018hfn}.

Sectioning of ${\cal A}$-theory  has been discussed in detail \cite{Hatsuda:2023dwx,Hatsuda:2024kuw}:
The Virasoro constraint is used to reduce 
spacetime only, while the  Gauss law constraint is used to reduce  both the spacetime and the worldvolume simultaneously \cite{Linch:2015fya,Linch:2015qva,Siegel:2018puf,Siegel:2020qef}.
{The realization of U-duality on the M2-brane has been extensively studied from various perspectives \cite{Duff:2015jka}, the M2-brane  has been shown to arise from the Virasoro dimensional reduction of the ${\cal A}$-theory brane \cite{Hatsuda:2024kuw}.
Dynkin diagrams of duality manifest theories show that
 ${\cal A}$-theory branes reduce to strings in D-dimensional spacetime
through a sequence of sectionings.}
The sectioning implemented by the Gauss law constraint introduces a new mechanism in which  choosing a solution of the Gauss law dimensional reduction condition reduces both the worldvolume and the spacetime. In this paper we analyze the general solutions of the Gauss law dimensional reduction condition and further investigate the exceptional symmetry covariant string solution. 
The exceptional $\sigma$-model has been widely studied \cite{Sakatani:2016sko,Sakatani:2017vbd,Sakatani:2017xcn,Arvanitakis:2017hwb,Arvanitakis:2018hfn,Blair:2019tww,
Sakatani:2022auu,Blair:2023noj,Osten:2024mjt}, and its relation to ${\cal A}$-theory provides a rich and nontrivial structure. In this paper, we propose this relation through the Gauss law constraint.

\par
\vskip 6mm
\subsection{Summary}

In this paper, we clarify the origin of the Gauss law constraint in the ${\cal A}$-theory brane and analyze its general solutions, demonstrating that the string solution is the unique consistent one
for lower dimensions. We then propose an exceptional symmetry covariant string solution that manifests an underlying two dimensional conformal symmetry.

In Section \ref{section:2}, we show that representations of spacetime, worldvolume, and the gauge parameter space in ${\cal A}$-theory correspond to the Dynkin roots of the exceptional group \cite{Linch:2016ipx}. 
The worldvolume gauge transformation of the spacetime coordinate is generated by the Gauss law constraint, and the symmetry transformation rule is
given by the CGW coefficient of the exceptional group.

In Section \ref{section:3}, we derive the Gauss law constraint from the Virasoro algebra of the ${\cal A}$-theory brane in D-dimensional space. We then present explicit expressions for the current algebra, the Virasoro algebra, and the Gauss law constraint in the general dimension case together with the explicit cases D=3 and D=4.

In Section \ref{section:4}, we examine solutions to the Gauss law dimensional reduction condition. We show that only the string solutions are consistent for D=3 and D=4. For higher dimensions  D$\geq$5 the additional constraint ${\cal V}=0$ governs the solution. 

In Section \ref{section:5}, we propose the exceptional symmetry covariant string solution.
 We also clarify its relation to the constant charge parameter in the exceptional $\sigma$-model \cite{Arvanitakis:2017hwb,Arvanitakis:2018hfn}.
 In this covariantized string solution the Virasoro algebra for a brane reduces to the standard Virasoro algebra for a string, showing that ${\cal A}$-theory branes exhibit precisely the two-dimensional conformal symmetry. This property is essential for the quantization of ${\cal A}$-theory.

\par
\vskip 6mm
\subsection{Conventions}
We use the following index conventions unless otherwise noted. 
\begin{enumerate}
	\item D denotes the spacetime dimension of the corresponding string theory; d denotes the worldvolume dimension. Both include time.
    \item $M, N, L, \cdots$ denote curved spacetime indices in general dimensions{;} $A, B, C, \cdots$ denote the corresponding flat spacetime indices.
	\item $m, n, p, \cdots$ denote curved worldvolume indices in general dimensions{;}  $a, b, c, \cdots$ denote corresponding flat worldvolume indices.
	\item ${\cal M}, {\cal N}, {\cal L}, \cdots$ denote indices of gauge parameters in general dimensions.    
    \item In D=4 case, $\mu, \nu, \rho, \cdots$ denote  16-component curved  spacetime spinor indices{;}  $\alpha, \beta, \gamma, \cdots$ denote corresponding flat spacetime indices.
\end{enumerate}

\par
\vskip 6mm
\section{Representations of coordinates}
\label{section:2}

\begin{figure}
[htbp]
  \centering
 \includegraphics[width=0.5\linewidth, trim=2mm 40mm 2mm 40mm ]{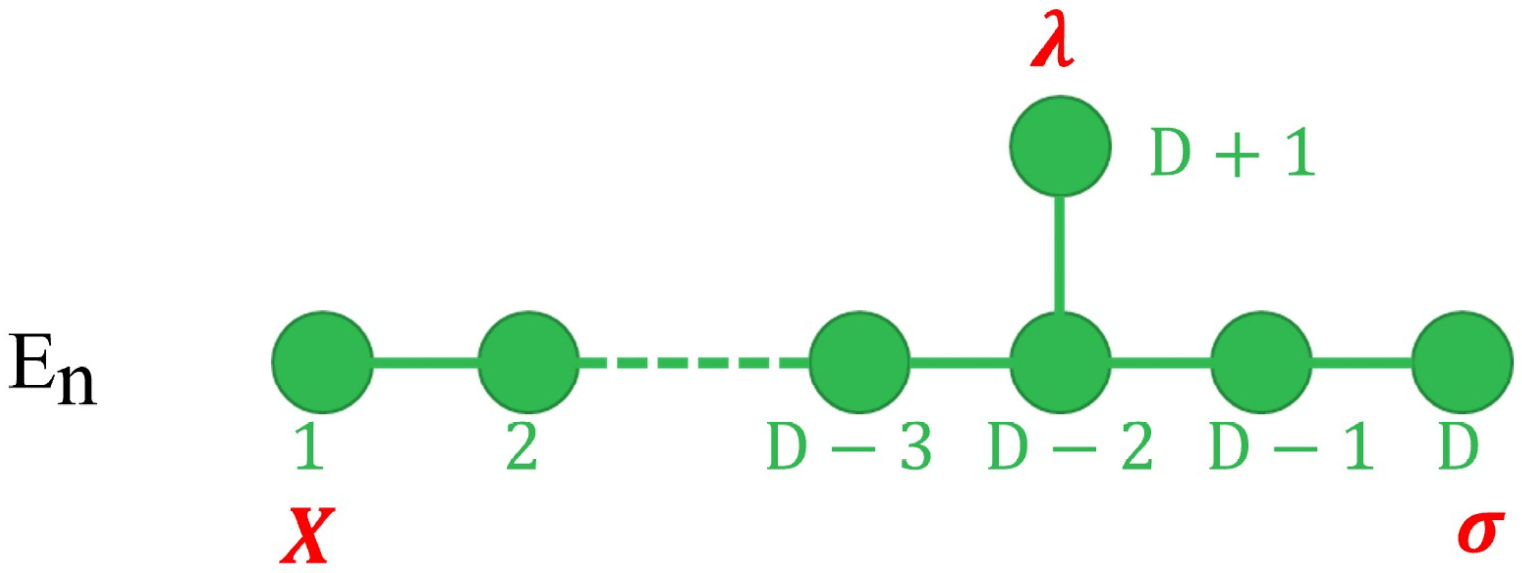}
  \caption{Representations of spacetime \Red{$X$}, worldvolume \Red{$\sigma$} and gauge parameter \Red{$\lambda$} for ${\cal A}$-theory corresponding  Dynkin node}
  \label{fig1:rep}
\end{figure}

In ${\cal A}$-theory the spacetime coordinate $X^M(\sigma_m)$ is promoted to a worldvolume gauge field, whose gauge transformation parameter is $\lambda^{\mathcal M}$.
The spacetime coordinate  \Red{$X$},  the brane worldvolume coordinate  \Red{$\sigma$} and the gauge parameter \Red{$\lambda$} are different representations of the exceptional group
as indicated by the Dynkin nodes in the Fig. \ref{fig1:rep}. 
{The representations of $X$, $\sigma$, $\lambda$ correspond to the particle, string, membrane multiplets in \cite{Obers:1998fb}, and 
$R_1$, $R_2$, $R_3$ in \cite{Berman:2012uy}.}
The worldvolume gauge transformation rule is described with the
CGW coefficient of the exceptional group 
${\mathbbm h}_{m{\cal M}}^M$  
as 
$\delta_\lambda X^M={\mathbbm h}^M_{m{\cal M}}\partial^m\lambda^{\cal M}$ 
\cite{Siegel:2018puf,Siegel:2019wrr}. 

\begin{figure}
[htbp]
  \centering
\includegraphics[width=0.8\linewidth, clip, trim=2mm 10mm 2mm 20mm ]{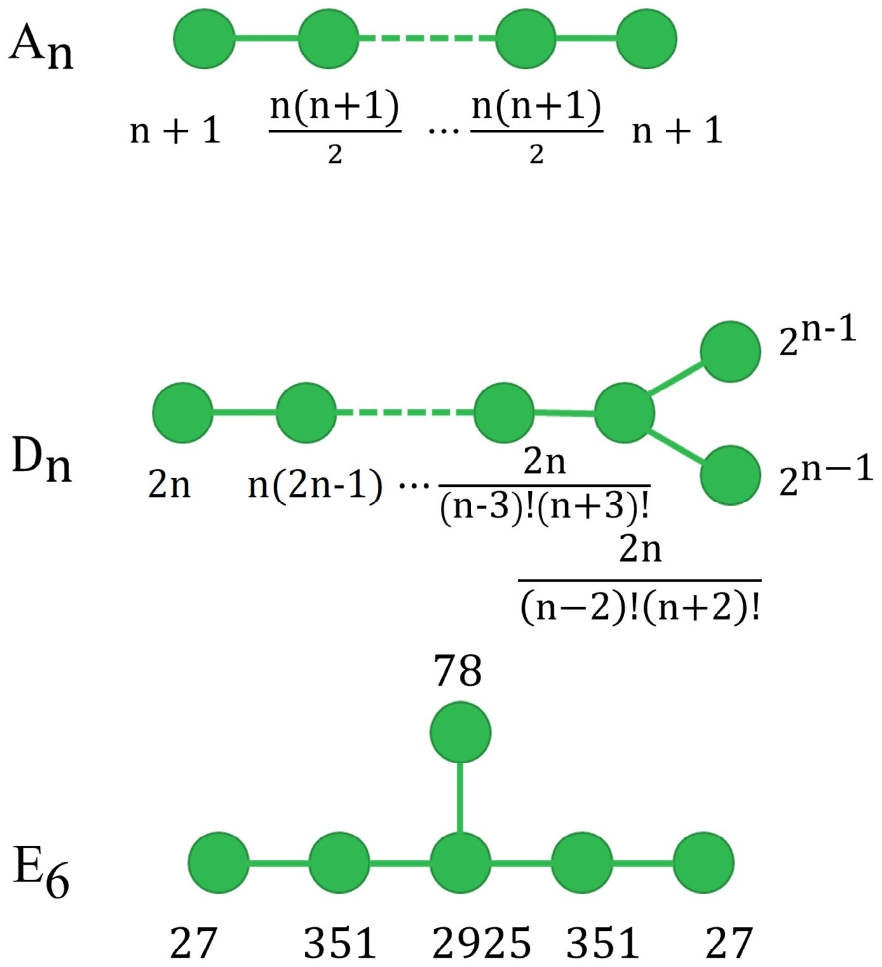}
  \caption{Dimensions of the fundamental representation with each Dynkin node}
  \label{fig2:rep}
\end{figure}

\begin{figure}
[htbp]
  \centering
 \includegraphics[width=0.55
 \linewidth,  trim=2mm 20mm 2mm 20mm ]{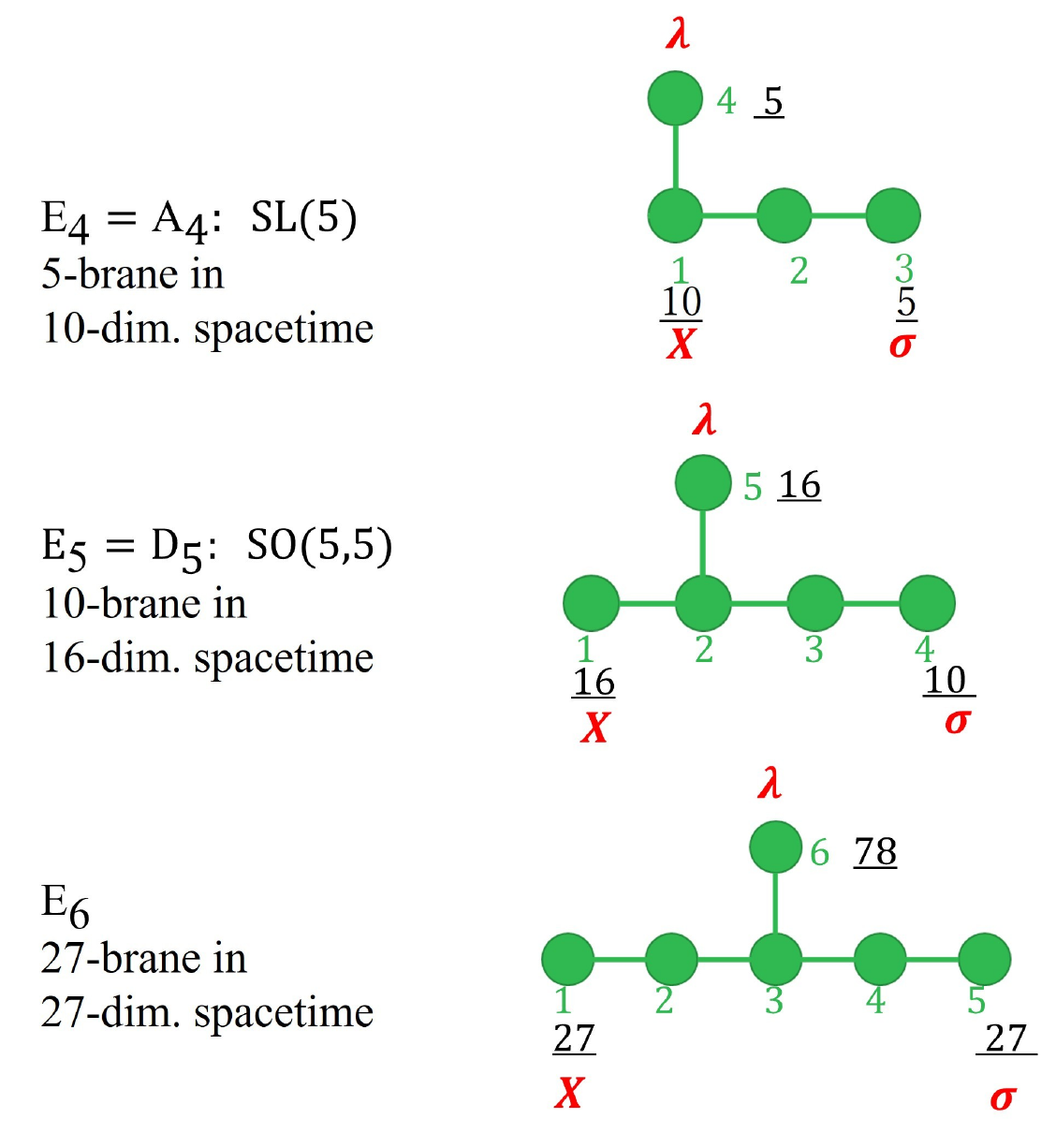}
 \vspace{3mm}
  \caption{Representations of spacetime coordinate \Red{$X$}, worldvolume coordinate \Red{$\sigma$} and gauge parameter \Red{$\lambda$} for  D=3, 4, 5 ${\cal A}$-theory}
  \label{fig3:rep}
\end{figure}

The space time coordinate, world volume coordinate,  and gauge parameter transform in the fundamental representations denoted in the above Dynkin nodes.
Dimensions of the fundamental representations associated with each Dynkin node of ${\rm A}_{\rm n}$, ${\rm D}_{\rm n}$ and ${\rm E}_6$ are 
given in Fig. \ref{fig2:rep} (for example \cite{Yamatsu:2015npn}).
Then dimensions of  spacetime, worldvolume and gauge parameter for
D=3, 4, 5  ${\cal A}$-theories 
are  shown in the Fig. \ref{fig3:rep}.

Corresponding representations of the spacetime coordinate, the worldvolume coordinate and the gauge parameter together with their dimensions are as follows.
\bea
{\renewcommand{\arraystretch}{1.3}
	\begin{array}{|c|c|cc|cc|cc|}
    \hline
		{\rm D}&G{\text -} {\rm symmetry}&
		{\rm spacetime}&X^M&{\rm worldvolume}&\sigma_m&{\rm gauge}&\lambda^{\mathcal M}\\\hline
		3&{\rm SL}(5)&
		\underline{10}&X^{mn}&\underline{5}&\sigma_m&\underline{5}&\lambda^m\\
		4&{\rm SO}(5,5)&
		\underline{16}&X^{\mu}&\underline{10}&\sigma_m&\underline{16}&\lambda^\mu\\
		5&{\rm E}_6&
		\underline{27}&X^M&\underline{27}&
		\sigma_m&\underline{78}&\lambda^{\mathcal M} \\\hline
\end{array}}
\eea 

\par
\vskip 6mm
\section{Gauss law constraint}\label{section:3}

\par  
\subsection{Origin of the Gauss law constraint}\par

 {We take the O(D,D) covariant current algebra as the starting point for } 
 ${\cal T}$-theory  { in the Hamiltonian formulation} 
\bea
\left[\dd_M(\sigma),\dd_N(\sigma')\right]&=&2i\eta_{MN}\partial_\sigma \delta(\sigma-\sigma')~~~
\eea
with the spacetime O(D,D) covariant derivative $\dd_M(\sigma)$.
 {$\eta_{MN}$ is}
the O(D,D) invariant metric
 {with the worldvolume index $\sigma$ suppressed, explicitly it is}
$\eta_{MN\sigma}$, and it  satisfies the   following property  
\bea
\eta^{MN}\eta_{NL}=\delta^M_L~~.\label{stringee}
\eea
The selfdual and anti-selfdual currents are given as
\bea
\dd_{M}&=&P_M+\eta_{MN}\partial_\sigma X^N\nn\\
\tilde{\dd}_{M}&=&P_M-\eta_{MN}\partial_\sigma X^N~~~\label{SDASD}
\eea
for the spacetime canonical coordinates $X^M(\sigma)$ and $P_M(\sigma)$.  {These currents \bref{SDASD} also correspond to the left and right moving currents, both of which are required to construct a worldsheet covariant O(D,D) Lagrangian.}
The selfduality condition is $\tilde{\dd}_M=0$.
 {The  Virasoro constraints ${\cal S}$ and ${\cal H}$ generate  diffeomorphisms in $\sigma$ and $\tau$ directions
using $H$ invariant metric $\hat{\eta}^{MN}$ on the coset space O(D,D)/$H$}
\bea
{\cal S}&=&\frac{1}{4}\dd_M\eta^{MN}\dd_N~=~0~~~\label{stringVcon}\\
 {
{\cal H}}&=& {\frac{1}{4}\dd_M\hat{\eta}^{MN}\dd_N~=~0~~~.}
\eea
Using  \bref{stringee}, they satisfy the closed Virasoro algebra
 {\bea
\left[ {\cal S}(\sigma), {\cal S}(\sigma')\right]
&=&{i}\left({\cal S}(\sigma)+{\cal S}(\sigma')\right)\partial_\sigma\delta(\sigma-\sigma')~~~\label{stringVirasoro}
\\
\left[ {\cal S}(\sigma), {\cal H}(\sigma')\right]&=&{i}({\cal H}(\sigma)+{\cal H}(\sigma'))\partial_\sigma\delta(\sigma-\sigma')~~~\nn\\
\left[ {\cal H}(\sigma), {\cal H}(\sigma')\right]&=&{i}({\cal S}(\sigma)+{\cal S}(\sigma'))\partial_\sigma\delta(\sigma-\sigma')~~~.\nn
\eea
}

 {We next extend the analysis to ${\cal A}$-theory. We take the E$_{\rm D+1}$ covariant brane current algebra as the starting point for } 
 ${\cal A}$-theory  { in the Hamiltonian formulation} 
\bea
\left[\dd_M(\sigma),\dd_N(\sigma')\right]&=&2i\eta_{MNk}\partial^k \delta(\sigma-\sigma')~~~\label{CABrane}
\eea
with the spacetime exceptional group covariant derivative $\dd_M(\sigma)$, 
the exceptional group invariant metric $\eta_{MNk}$,
  the (d$-1$)-brane worldvolume derivative $\partial^k \delta(\sigma-\sigma')
  =\displaystyle\frac{\partial}{\partial \sigma_k}\delta^{(\rm d-1)}(\sigma-\sigma') $.
In contrast to the string case \bref{stringee}
 the exceptional group invariant metric $\eta^{MNl}$ 
has the following property
\bea
\eta^{MLm}\eta_{NLn}&=&\delta^M_N\delta^m_n-U^{Mm}_{Nn}~~\label{braneee},~~
U^{Mm}_{Nn}
~=~{\mathbbm h}^{M}_{n {\cal M}}
 {\mathbbm h}_{N}^{m {\cal M}}
\eea
with $\eta^{MNm}\eta_{LNm}=\delta^M_L$.
 {It should be noted that, for E$_{D+1}$ with D$\geq 5$, Eq. (3.7) holds only weakly, subject to the additional constraint ${\cal V}=0$. For the E$_6$ case, see Ref.\cite{Siegel:2018puf} .}
The selfdual and anti-selfdual  {(the left/right moving)} currents  are given as
\bea
\dd_{M}&=&P_M+\eta_{MNm}\partial^m X^N\nn\\
\tilde{\dd}_{M}&=&P_M-\eta_{MNm}\partial^m X^N~~~,
\eea
where the selfduality condition is $\tilde{\dd}_M=0$.
The brane Virasoro constraint  {${\cal S}^m$  generates diffeomorphisms in spatial directions of the brane worldvolume $\sigma$ }
\bea
{\cal S}^m&=&\frac{1}{4}\dd_M\eta^{MNm}\dd_N~=~0 ~~~,\label{BraneVirasoro}
\eea
 {while Virasoro constraint 
 ${\cal H}$ generates the diffeomorphism in $\tau$ direction
 using the $H$ invariant metric $\hat{\eta}^{MN}$ on the coset space $G/H$ }
\bea
 {{\cal H}}&=& {\frac{1}{4}\dd_M\hat{\eta}^{MN}\dd_N~=~0 ~~~.} 
\eea

They satisfy the following algebra with \bref{braneee}
\bea
\left[{\cal S}^m(\sigma),{\cal S}^n(\sigma')\right]&=&
\frac{i}{2}\eta^{M_1M_2m}\eta_{M_2 N_2 l}\eta^{N_1N_2n}
\dd_{M_1}(\sigma)\dd_{N_1}(\sigma')\partial^l\delta(\sigma-\sigma')
\nn\\
&=&i
\left({\cal S}^{(m}(\sigma)\partial^{n)}\delta(\sigma-\sigma')
+{\cal S}^{(m}(\sigma')\partial^{n)}\delta(\sigma-\sigma')~\right)\nn\\
&&-i\eta^{KL(m}{\mathbbm h}^{n){\cal M}}_{L}{\mathbbm h}_{l{\cal M}}^M
\left(\dd_K\dd_M(\sigma)+\dd_K\dd_M(\sigma')\right)\partial^l\delta(\sigma-\sigma')\nn\\
&&-i\left({\mathbbm h}^{[m|{\cal M}}_L\eta^{LN|n]}{\mathbbm h}^{M}_{l{\cal M}}\dd_M\partial^l\dd_N
-{\mathbbm h}^{[m|{\cal M}}_L\eta^{LM|n]}{\mathbbm h}^{N}_{l{\cal M}}\dd_M\partial^l\dd_N
\right)\nn\\&&~~~~~~~~~~~~~~~~~~~~~~~~~~~~~~~~~~~~~~~~~~~~~~~~~~~~~~~~~~~~~~~\times\delta(\sigma-\sigma')\nn\\
&=&i
\left({\cal S}^{(m}(\sigma)\partial^{n)}\delta(\sigma-\sigma')
+{\cal S}^{(m}(\sigma')\partial^{n)}\delta(\sigma-\sigma')~\right)\nn\\
&&-i\eta^{KL(m}{\mathbbm h}^{n){\cal M}}_{L}\left(\dd_K\vec{\cal U}_{\cal M}(\sigma)
-\dd_K\vec{\cal U}_{\cal M} (\sigma')\right)\delta(\sigma-\sigma')\nn\\
&&-i\delta(\sigma-\sigma')
\left({\mathbbm h}^{[m|{\cal M}}_L\eta^{LN|n]}
\vec{\cal U}_{{\cal M}}\dd_N
-{\mathbbm h}^{[m|{\cal M}}_L\eta^{LM|n]}\dd_M
{\cal U}_{{\cal M}}
\right)~~\label{SSSU}
~~.
\\
 {\left[{\cal S}^m(\sigma),{\cal H}(\sigma')\right]}&=&
 {i\left({\cal H}(\sigma)+{\cal H}(\sigma')\right)
\partial^{m)}\delta(\sigma-\sigma')}\nn\\
&& {-\frac{i}{2}\left(
\hat{\eta}^{KL}
{\mathbbm h}^{m{\cal M}}_{L}
\dd_K\vec{\cal U}_{\cal M}(\sigma)\delta(\sigma-\sigma')
+\delta(\sigma-\sigma')(\vec{\cal U}_{\cal M} \dd_K) \right) }\nn\\
 {\left[{\cal H}(\sigma),{\cal H}(\sigma')\right]}&=&
 {i
\left({\cal S}^m(\sigma)+{\cal S}^m(\sigma')\right)
\partial_{m}\delta(\sigma-\sigma')}\label{SH}~~.
\eea
In the second line from the bottom of \bref{SSSU}   $\partial^m\delta(\sigma-\sigma')=\frac{\partial}{\partial \sigma_m}\delta(\sigma-\sigma')=-\frac{\partial}{\partial \sigma'_m}\delta(\sigma-\sigma')$ is used.
The closure of the Virasoro algebra requires the Gauss law constraint
\bea
{\cal U}_{\cal M} &=&{\mathbbm h}^{M}_{n{\cal M}}\partial^n\dd_M=0~~~,\label{Gausslaw}
\eea
which also acts on an arbitrary function $f(\sigma)$ as
\bea
~~\vec{\cal U}_{\cal M}f(\sigma)={\mathbbm h}^{M}_{n{\cal M}}\dd_M\partial^n f(\sigma)=0~~~.\label{sectionU}
\eea
The Gauss law constraint is imposed both as a strong condition and as a weak condition on the zero modes of the spacetime coordinates and the worldvolume coordinates. Explicitly it is written as
\bea
{\cal U}_{\cal M} f~=~
{\mathbbm h}^{M}_{n{\cal M}}\partial^n\left(\frac{1}{i}\partial_M f\right)~=~0~~,~~{\cal U}_{\cal M}(f,g)~=~
{\mathbbm h}^{M}_{n{\cal M}}(\partial^n f)\left(\frac{1}{i}\partial_M g\right)~=~0 \label{Gausssection}
\eea
for arbitrary functions $f(X(\sigma))$ and $g(X(\sigma))$.

The worldvolume gauge transformation is generated by the Gauss law constraint as
\bea
\delta_\lambda X^M &=&\left[\displaystyle\int \frac{1}{i}\lambda^{\cal M}
{\cal U}_{\cal M}, X^{M}(\sigma)\right] ~=
~{\mathbbm h}^{M}_{n{\cal M}}\partial^n\lambda^{\cal M} ~~~.
\eea

 {The relation of this Gauss law constraint to the one for the U(1) gauge fields on D-brane worldvolumes is an interesting question. Just as the massless excitations of an open string with the Neumann directions give rise to the U(1) gauge field on a D-brane, 
one may expect an analogous mechanism when an appropriate section of an A-theory brane reproduces a D-brane. 
If the degrees of freedom corresponding to the U(1) gauge field can be identified within this section, 
the Gauss law constraint discussed above may reduce to the Gauss law for the worldvolume U(1) gauge field. 
At present, however, the D-brane section of ${\cal A}$-theory has not yet been fully understood, so we leave this issue for future investigation.}

\par
\vskip 6mm
\subsection{D=3: SL(5) case}

In D=3  ${\cal A}$-theory enjoys ${\rm SL}(5)$ U-duality, with invariant metric 
$\eta^{MNm}=\epsilon^{m_1m_2n_1n_2m}$  with
$m=1,\cdots,5$. 
The SL(5) current algebra is given by
\bea
\left[\dd_{m_1m_2}(\sigma),\dd_{m_3m_4}(\sigma')\right]&=&
2i\epsilon_{m_1\cdots m_4 m}\partial^{m}\delta(\sigma-\sigma')~~~.
\eea
Here $\delta(\sigma-\sigma')$ denotes the delta function over the 5-dimensional worldvolume $\delta^{(5)}(\sigma-\sigma')$.
The group invariant metric $\eta^{MNm}=\epsilon^{m_1m_2n_1n_2 m}$ satisfies the following relations
\bea
\eta^{MLm}\eta_{NLn}&=&\frac{1}{2}\epsilon^{m_1m_2l_1l_2 m}\epsilon_{n_1n_2l_1l_2 n}\nn\\
&=&\delta^{m_1}_{[n_1}\delta^{m_2}_{n_2}\delta^{m}_{n]}\nn\\
&=&\delta^{m_1}_{[n_1}\delta^{m_2}_{n_2]}\delta^{m}_{n}
+\frac{1}{2}\delta^{[m_1}_{n}\delta^{m_2]}_{[n_1}\delta^{m}_{n_2]}
+\frac{1}{2}\delta^{[m_1}_{[n_2|}\delta^{m_2]}_{n}\delta^{m}_{|n_1]}\nn\\
&=&\delta^M_N\delta^m_n-U^{Mm}_{Nn}\nn\\
U^{Mm}_{Nn}&=&\delta^{[m_1}_n\delta^{m_2]}_{l}
\delta^{m}_{[n_1}\delta^{l}_{n_2]}
~=~{\mathbbm h}^M_{nl}{\mathbbm h}_N^{ml}
\label{U}\nn\\
{\mathbbm h}^M_{n{\cal M}}&=&\delta^{[m_1}_{n}\delta^{m_2]}_{l}\label{CGWh}~~~.
\eea
The Virasoro constraint and the Gauss law constraint are given by
\bea
{\cal S}^m&=&\frac{1}{16}\epsilon^{m_1m_2n_1n_2m}\dd_{m_1m_2}\dd_{n_1n_2}\label{ViraD=3}\\
{\cal U}_m&=&\partial^n\dd_{mn}\label{GaussD=3}~~~.
\eea
The Virasoro algebra with the Gauss law constraint for D=3 SL(5) ${\cal A}$-theory is given by \cite{Hatsuda:2023dwx}.

The SL(5) current is given by
\bea
\dd_{m_1m_2}&=&P_{m_1m_2}+\frac{1}{2}\epsilon_{m_1\cdots m_5}\partial^{m_3}X^{m_4m_5}~~~.
\eea
The spacetime coordinate is the SL(5) rank two antisymmetric tensor $X^{mn}$ and the 5-brane worldvolume coordinate is
 the SL(5) vector $\sigma_m$ where they  satisfy the following canonical commutators
\bea
\left[P_{mn}(\sigma),X^{lk}(\sigma')\right]&=&\frac{1}{i}\delta^{[l}_m\delta^{k]}_n\delta(\sigma-\sigma')~~~\nn\\
\left[\partial^m,\sigma_n\right]&=&\delta^{m}_n~~~.
\eea

The worldvolume gauge transformation is given by
\bea
\delta_\lambda X^{m_1m_2}&=&\left[\frac{1}{i}\displaystyle\int \lambda^n{\cal U}_n, X^{m_1m_2}\right]\nn\\&=&
{\mathbbm h}^M_{ln}\partial^l\lambda^{n}~=~
\delta_l^{[m_1}\delta_{n}^{m_2]}~\partial^l\lambda^{n}
~=~\partial^{[m_1}\lambda^{m_2]}~~~.
\eea

\par
\vskip 6mm
\subsection{D=4: SO(5,5) case}

In D=4  ${\cal A}$-theory with SO(5,5) U-duality the group invariant metric is  
$\eta^{MNm}=(C\gamma^m){}^{\mu \nu}$  with
$m=1,\cdots,10$,  $\mu,~\nu=1,\cdots, 16$ and the charge conjugation matrix $C$. 
The SO(5,5) current algebra is given by
\bea
\left[\dd_{\mu}(\sigma),\dd_{\nu}(\sigma')\right]&=&
2i(C\gamma_m)_{\mu\nu }\partial^{m}\delta(\sigma-\sigma')~~~,
\eea
with  $\left\{\gamma_m,\gamma_n\right\}=2\hat{\eta}_{mn}$ 
for the SO(5,5) metric  
$\hat{\eta}_{mn}=(1,\cdots,1;-1,\cdots,-1)$.
It refers to the delta function over the 10-dimensional worldvolume
$\delta(\sigma-\sigma')=\delta^{(10)}(\sigma-\sigma')$.
The group invariant metric 
$\eta_{MNm}=\frac{1}{\sqrt{2}}(C\gamma_m)_{\mu\nu}$  satisfies the following relations
\bea
\eta^{MLm}\eta_{NLn}&=&\frac{1}{2}(\gamma^m C^{-1})^{\mu\lambda }(C\gamma_n)_{\lambda \nu }\nn\\
&=&\delta_{n}^{m}\delta_{\nu}^{\mu}-\frac{1}{2}\gamma_{n}{}^{\mu}{}_{\lambda}\gamma^m{}^{\lambda}{}_{ \nu}
\nn\\
U^{Mm}_{Nn}&=&\frac{1}{2}\gamma_n{}^{\mu}{}_{\lambda}  \gamma^m{}^\lambda{}_{ \nu}=
{\mathbbm h}^{\mu}_{n\lambda} {\mathbbm h}_{\nu}^{m\lambda} \nn\\
{\mathbbm h}^M_{n{\cal M}}&=&\frac{1}{\sqrt{2}} \gamma_n{}^{\mu}{}_{\lambda}~~~
\eea
with the 16-dimensional representation ${\cal M}=\lambda$. 

For D$\geq$4 cases the closure of the Virasoro constraint and the Gauss law constraint requires the further constraint ${\cal V}=0$ 
which is the product of the two worldvolume derivatives contracted with 
the Clebsch–Gordan-Wigner (CGW) coefficients.
For D=4 case, the Virasoro constraint and the Gauss law constraint together with the additional constraint
${\cal V}=0$ are given by
\bea
{\cal S}^m&=&\frac{1}{4}(C\gamma^m)^{\mu\nu }\dd_\mu\dd_\nu~=~0\nn\\
{\cal U}_\mu&=&\gamma_n{}^{\nu}{}_{\mu} \partial^n\dd_{\nu}~=~0~\\
{\cal V}&=&\hat{\eta}_{mn}\partial^m \partial^n~=~0~~~.\nn
\eea
The Virasoro algebra with the Gauss law constraint for D=4 SO(5,5) 
${\cal A}$-theory is  given by \cite{Hatsuda:2021wpb}.

The SO(5,5) current is given by
\bea
\dd_\mu&=&P_\mu+\partial^m X^\nu(C\gamma_m)_{\nu\mu}~=~
P+ \bar{X} {\slashed{\partial}}
~,
\eea
where the worldvolume derivative operator $\slashed{\partial}=\partial^m \gamma_{m}$ acts on $X^\mu$
 {with $\bar{X}=XC$}.
The spacetime coordinate is the SO(5,5) spinor  $X^{\mu}$, $\mu=1,\cdots, 16$, and the worldvolume coordinate is the SO(5,5) vector $\sigma^m$, $m=1,\cdots 10$ which satisfy the following canonical commutators
\bea
\left[P_{\mu}(\sigma),X^{\nu}(\sigma')\right]&=&\frac{1}{i}\delta^{\nu}_{\mu}\delta(\sigma-\sigma')\label{PandXmu}\\
\left[\partial^m,\sigma_n\right]&=&\delta^{m}_n~~~.\nn
\eea

The Gauss law constraint  is
\bea
{\cal U}_\mu&=&~P\slashed{\partial}+{\cal V} \bar{X} ~\approx~
P\slashed{\partial}~=~0~~~.
\eea
The Gauss law constraint generates the worldvolume gauge transformation rule of the spacetime coordinate as
\bea
\delta_\lambda X^{\mu}&=&
\gamma_n{}^\mu{}_\nu\partial^n\lambda^{\nu}
~=~\slashed{\partial}\lambda~~~.
\eea

\par
\vskip 6mm
\section{Solutions of the Gauss law dimensional reduction condition}\label{section:4}

\subsection{Sectionings}

It is well known that, in the 11-dimensional supergravity theory as the low energy limit of the M-theory, a membrane reduces to a string through double dimensional reduction, where one dimension is simultaneously compactified in both the spacetime and the worldvolume.
This simultaneous reduction of spacetime and worldvolume is generalized within ${\cal A}$-theory.

\begin{figure}[h]
	\centering
 \includegraphics[width=0.9\linewidth, 
trim=2mm 2mm 2mm 2mm]{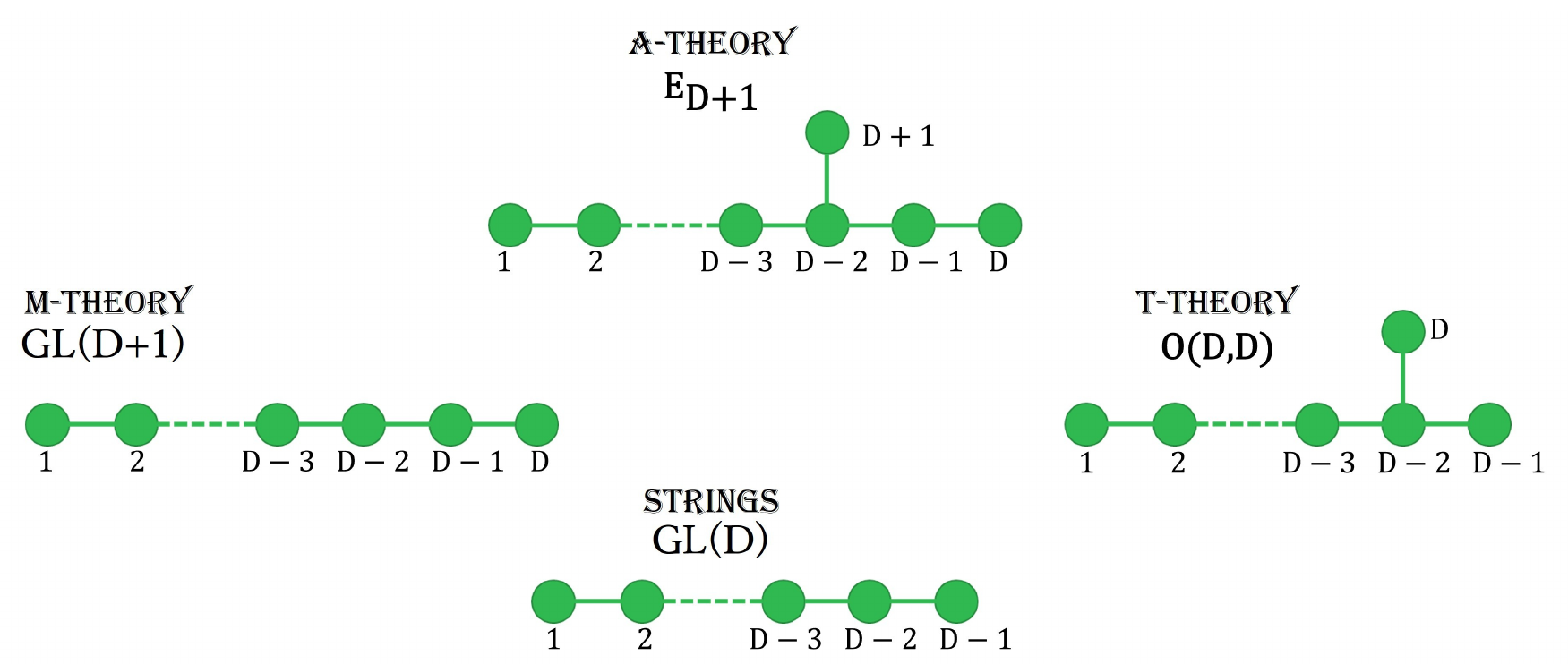}
	\caption{Dynkin diagrams of duality manifest theories}
	\label{Diamond}
\end{figure}

Dynkin diagrams of duality manifest theories are 
 shown in Fig. \ref{Diamond}.
These theories are described by duality symmetry covariant branes; 
exceptional  ${\cal A}$-theory brane,  ${\cal M}$-theory brane,  ${\cal T}$-string.

The Virasoro constraint and the Gauss law constraint are background independent.
 {For example, defining curved space current $\dd_A=E_A{}^M\dd_M$ and 
using the identity     $E_A{}^M\eta^{ABc}E_B{}^N$
$=\eta^{MNm}E_m{}^c$ one obtains 
${\cal S}^c=\dd_A\eta^{ABc}\dd_B=E_m{}^{c}{\cal S}^m$.
Therefore $S^m=0$ remains valid even in a curved background.}
Therefore they are used to reduce either the spacetime or the worldvolume, or both.
Replacing spacetime coordinates with their 0-modes leads to  conditions for sectioning \cite{Kugo:1992md,Linch:2015fya,Linch:2015qva}. 
The section condition based on the Virasoro constraint  is the condition for the effective low energy gravitational theories such as DFT \cite{Siegel:1993bj,Siegel:1993xq,Siegel:1993th,Hohm:2010pp} and  ExFT \cite{Berman:2011cg,Coimbra:2011ky}.
In ${\cal A}$-theory the section condition based on the Virasoro constraint is bilinear in the 0-mode of the spacetime momentum $p_M=\frac{\partial}{i\partial x^M}$, while the section condition based on the Gauss law constraint is a product of the 0-mode of   the spacetime momentum and worldvolume derivative $\partial^m=\frac{\partial}{\partial \sigma_m}$.

On the other hand the dimensional reduction condition is linear in the spacetime momentum $P_M(\sigma)$ including all massive modes. 
 {The dimensional reduction condition
$\widetilde{P}_L(\sigma)-\widetilde{P}_R(\sigma)=0$
was introduced in Eq. (64) of  \cite{Polacek:2013nla}. 
Since the anti-selfdual (right moving) current commutes with the selfdual (left moving or covariant derivative) current, this condition is first class. Its gauge fixing,
$\widetilde{X}(\sigma)=X_L(\sigma)-X_R(\sigma)=0$, 
eliminates the unphysical degrees of freedom at both the massless and massive levels, as explained in  detail \cite{Hatsuda:2015cia}.}
The dimensional reduction conditions for brane worldvolume theories are given in \cite{Linch:2015fya,Linch:2015qva}. 
 {The E$_{D+1}$ covariant dimensional reduction conditions are linear in the momentum with full massive modes  multiplied with the zero mode of the momentum as the coefficient, in the form of the Virasoro constraint $p\eta P(\sigma)=0$ and in the form of the Gauss law constraint $\partial {\mathbbm h} P(\sigma)=0$.}

Exceptional covariant dimensional reduction conditions and section conditions are listed below.
For higher dimensions, the additional constraint ${\cal V}=0$ reduces the worldvolume dimension.
\bea
{\renewcommand{\arraystretch}{1.8}
	\begin{array}{llll}
		{\rm Virasoro~and~Gauss}& {\cal S}^m=\frac{1}{4}\eta^{MNm}\dd_{M}\dd_{N}&
		{\cal U}_{\cal M}={\mathbbm h}_{n{\cal M}}^M\partial^n\dd_{M}
	&  		\\
		{\rm Dimensional~reduction}& \overline{\cal S}^m=\frac{1}{2}\eta^{MNm}p_{M}P_{N}(\sigma)&
		\overline{\cal U}_{\cal M}={\mathbbm h}_{n{\cal M}}^M\partial^nP_{M}(\sigma)
		&
		\\
		{\rm Section~condition}& \underline{\cal S}^m=\frac{1}{4}\eta^{MNm}p_{M}p_{N}&
		\underline{\cal U}_{\cal M}={\mathbbm h}_{n{\cal M}}^M\partial^n p_{M}&
	{\cal V}=\partial\partial
\end{array}}
\eea

Reductions of spacetime,
 ${\cal A}$ $\to$ ${\cal M}$-theories and ${\cal T}$ $\to$ S-theories, are realized by solving 
  the Virasoro dimensional reduction condition  
   $\overline{S}=0$.
Reductions of worldvolume, 
${\cal A}$ $\to$ ${\cal T}$-theories and ${\cal M}$ $\to$ S-theories, are realized   by solving the Gauss law dimensional reduction condition  
$\overline{\cal U}=0$,
where some solutions require the reduction of the spacetime. The sectionings obtained through these dimensional reductions form a diamond shaped structure.
\bea
&\begin{array}{c}
	\begin{array}{c}
		\cal{A}\mathchar`-\rm{theory}\\
		{\rm brane~with~}	{\rm E}_{\rm D+1}
	\end{array}
	\\\\
	{\cal S}~	\swarrow\quad\quad\quad\quad\quad
	\searrow~{\cal U}\\\\
	\begin{array}{c}
		\cal{M}\mathchar`-\rm{theory}\\ 
		{\rm brane ~with~}{\rm GL}(D+1)
	\end{array}
	\quad\quad\quad\quad\quad\quad
	\begin{array}{c}
		\cal{T}\mathchar`-\rm{theory}\\
		{\rm string~with~}{\rm O(D,D)}
	\end{array}\\\\
	{\cal U}~	\searrow\quad\quad\quad\quad\quad\quad\quad\quad\swarrow~{\cal S}\\\\
	\begin{array}{c}
		\rm{S}\mathchar`-\rm{theory}\\{\rm string ~with~} {\rm GL(D)}
	\end{array}
\end{array}\label{AGATMSWeb}&\nn\\
\\&\rm{Figure ~5}~:{\rm Sectionings~of~} {\cal A}\mathchar`-{\rm theory ~diamond}
&\nn
\eea
In the ${\cal A}$-theory diamond diagram in \bref{AGATMSWeb}, the southeast pointing arrows terminate at the string worldsheet theory by solving the Gauss law dimensional reduction condition.
In this paper, we examine whether this is indeed the unique consistent solution or whether other brane solutions are also allowed.
\par
\vskip 6mm

\subsection{D=3: SL(5) case}

In the D=3 case both the worldvolume and spacetime coordinates carry tensor indices so sectioning is straightforward.
We first solve the Gauss law dimensional reduction condition  
to reduce the 5-brane worldvolume into a string worldsheet.
We then examine the consistency of the theories as duality covariant theories in the ${\cal A}$-theory diamond Fig. {\color{green}5} in \bref{AGATMSWeb} in the  membrane and  3-brane cases. 
For D=3 SL(5) case, a brane theory obtained from the 5-brane in the 10-dimensional spacetime by solving the Gauss law dimensional reduction condition must have the spacetime dimension larger than three.  {This is because the Dynkin diagram of Fig. \ref{Diamond} indicates that 
the Virasoro dimensional reduction constraint reduces to the string theory in 3 dimensions.}
We show that enlarging the worldvolume leads to a stronger restriction on the spacetime imposed by the Gauss law dimensional reduction condition, so that only the string solution is a consistent solution.
\begin{enumerate}
	\item
	{\bf String}
	
The Gauss law dimensional reduction condition is solved by the string worldsheet solution upon choosing $\sigma_5=\sigma$ 
\bea
\partial_\sigma&=&\partial^5~\neq~ 0~~,~~\partial^i~=~0
~~,~~i=1,\cdots,4\label{stringD3}\\
\overline{\cal U}_5&=&\partial^i P_{i5}~=~0~~,~~
\overline{\cal U}_i~=~\partial^5 P_{5i}+\partial^j P_{ji}~=~0\nn\\
\Rightarrow && P_{5i}~=~0 ~~,~~P~=~P_{ji}~=~-P_{ji}~~.
\label{ststringD3}
\eea
The spacetime coordinate is $X^{ij}$ with a spacetime dimension of 6. 
This is a ${\cal T}$-string in the O(3,3) spacetime.
The covariant derivative has the following components with $\epsilon_{ijkl}=\epsilon_{ijkl5}$
\bea
\dd_{ij}&=&P_{ij}+\frac{1}{2}\epsilon_{ijkl}\partial^5X^{kl}
~~,~~
\dd_{5i}~=~0~~~
\eea
 {where $\dd_{ij}$ corresponds to the selfdual string current \bref{SDASD}.}
The Virasoro operator has the following nontrivial component
\bea
{\cal S}^5&=&\frac{1}{16}\dd_{ij}\epsilon^{ijkl}\dd_{kl}~~,~~{\cal S}^{i}~=~0~~~
\eea
 {where ${\cal S}^5$ corresponds to the string Virasoro constraint \bref{stringVcon}.}
The Virasoro constraint ${\cal S}^5=0$ generates the diffeomorphism in the $\sigma$ direction.
The dimensional reduction condition for the Virasoro operator ${\cal S}^5$  allows the T-duality covariant spacetime 
	\bea
	\overline{\cal S}^5=\frac{1}{8}\epsilon^{ijkl}p_{ij}P_{kl}(\sigma)= 0
	\eea
	and its section condition 
	\bea
	\underline{\cal S}^5=\frac{1}{16}\epsilon^{ijkl}p_{ij}p_{kl} =0~~~
	\eea
selects the physical spacetime.
This string solution, \bref{stringD3} and \bref{ststringD3},
is 	a consistent solution of the Gauss law dimensional reduction condition representing the ${\cal T}$-string in O(3,3) spacetime.
The string in the 3-dimensional spacetime is obtained by the further sectioning \cite{Hatsuda:2024kuw}.
	
\item
{\bf Membrane}

	The Gauss law dimensional reduction condition is solved assuming the membrane worldvolume  directions  to be  
$\sigma_\alpha$ with $\alpha=4,5$ 
\bea
\partial^\alpha&\neq&0~~,~~\partial^i~=~0~~,~~\alpha=4,5~~,~~i=1, 2, 3\label{braneD3}\\
\overline{\cal U}_\alpha&=&\partial^i P_{i\alpha}+\partial^\beta P_{\beta\alpha}
~=~0~~,~~
\overline{\cal U}_i~=~\partial^\alpha P_{\alpha i}+\partial^j P_{ji}~=~0\nn\\
\Rightarrow && P_{\alpha i}~=~P_{\alpha \beta}~=~0 ~~,~~P~=~P_{ji}~~,~~j \neq i~~~.
\label{stbraneD3}
\eea
The spacetime coordinate is $X^{ij}$ with a spacetime dimension of 3. 
Since the spacetime dimension must be larger than 3, the membrane solution is not a suitable solution.
 {We regard the D=3 membrane solution as inconsistent because it does not fit the dimensional reduction structure of the diamond diagram. A consistent reduction of the SL(5) ${\cal A}$-theory brane should ultimately lead to the D=3 string. However, although the southeast path Gauss-law reduction may yield a D=3 membrane, the subsequent southwest path Virasoro reduction lowers only the spacetime dimension, not the worldvolume dimension. Therefore, there is no path that reduces the D=3 membrane to the D=3 string.}

The covariant derivative has the following components with $\epsilon_{\alpha  \beta}\epsilon_{ijk}=\epsilon_{\alpha  \beta ijk}$
\bea
\dd_{ij}&=&P_{ij}
~~,~~
\dd_{\alpha i}~=~\frac{1}{2}\epsilon_{\alpha  \beta}\epsilon_{ijk}\partial^\beta X^{jk}~~,\dd_{\alpha\beta}~=~0~~~.
\eea
The Virasoro operator has the following non-trivial component
\bea
{\cal S}^\alpha&=&\frac{1}{4}\dd_{\beta i}\epsilon^{\alpha\beta}\epsilon^{ijk}\dd_{jk}~=~-\frac{1}{4}(\partial^\alpha X^{ij})P_{ij}\nn\\
{\cal S}^{i}&=&\frac{1}{4}\epsilon_{\alpha \beta } \epsilon_{jkl}
(\partial^\alpha X^{ij})
(\partial^\beta X^{kl})
~~~,
\eea
which satisfy the closed algebra
	\bea
	\left[{\cal S}^\alpha(\sigma), {\cal S}^\beta(\sigma') \right]&=&i\left\{
    {\cal S}^{(\alpha}(\sigma)\partial^{\beta)}\delta(\sigma-\sigma')
    +{\cal S}^{(\alpha}(\sigma') \partial^{\beta)}\delta(\sigma-\sigma')\right\}\nn\\  
	&&-i\partial^{[\alpha}{\cal S}^{\beta]}\delta(\sigma-\sigma')\\
	\left[{\cal S}^\alpha(\sigma), {\cal S}^i(\sigma') \right]&=&-i\left\{
	\left({\cal S}^i(\sigma) +{\cal S}^i(\sigma')   \right)\partial^\alpha\delta(\sigma-\sigma')
	-\partial^\alpha {\cal S}^i \delta(\sigma-\sigma')\right\}\nn\\
	\left[{\cal S}^i(\sigma), {\cal S}^j(\sigma') \right]&=&0\nn
\eea
The Virasoro constraint ${\cal S}^\alpha=0$ generates the  reparametrization of the worldvolume coordinate $\sigma_\alpha$ 
\bea
\left[i\displaystyle\int \xi_\alpha{\cal S}^\alpha, X^{ij}(\sigma) \right]=\xi_\alpha \partial^\alpha X^{ij}~~.
\eea
However  ${\cal S}^i=0$ is the following relation with $~{\bf X}_i=\epsilon_{ijk}X^{jk}=(X,Y,Z)$
\bea
\left\{\begin{array}{ccl}
{\cal S}^X&=&\partial^1 Z \partial^2 Y-\partial^2 Z\partial^1 Y=0\nn\\
{\cal S}^Y&=&\partial^1 X \partial^2 Z-\partial^2 X\partial^1 Z=0\nn\\
{\cal S}^Z&=&\partial^1 Y\partial^2 X-\partial^2 Y\partial^1 X=0\nn
\end{array}\right.~\Leftrightarrow~~
\begin{array}{c}
{\bf S}~=~{\bf v}^1\times {\bf v}^2~=~0\\
{\bf v}^\alpha=\partial^\alpha {\bf X}
\end{array}~~~,
\eea
which implies that the membrane surface area collapses, or equivalently the local area density vanishes.
This indicates that the membrane solution of the Gauss law constraint is inconsistent.
The sectioning along the southwest arrow in the ${\cal A}$-theory diamond \bref{AGATMSWeb} cannot be consistently performed, since 
 both the Virasoro dimensional reduction condition 
and the section condition have no
non-trivial components
\bea
{\rm dimensional~reduction}&:&
\overline{\cal S}^\alpha~=~0~~~\nn\\
{\rm section~condition}&:&
\underline{\cal S}^\alpha~=~0~~~.
\eea 
Therefore it is not suitable as a duality manifest brane theory.

\item
{\bf 3-brane}

	The Gauss law dimensional reduction condition is solved assuming the 3-brane worldvolume  directions  to be  
$\sigma_\alpha$ with $\alpha=3,4,5$ 
\bea
\partial^\alpha&\neq&0~~,~~\partial^i~=~0~~,~~i=1, 2 \label{3braneD3}\\
\overline{\cal U}_\alpha&=&\partial^i P_{i\alpha}+\partial^\beta P_{\beta\alpha}
~=~0~~,~~
\overline{\cal U}_i~=~\partial^\alpha P_{\alpha i}+\partial^j P_{ji}~=~0\nn\\
\Rightarrow && P_{\alpha i}~=~P_{\alpha \beta}~=~0 ~~,~~P~=~P_{ji}~~,~~j \neq i~~~.
\label{3stbraneD3}
\eea
The spacetime coordinate is $X^{ij}$ with a spacetime dimension of 1. 
The 3-brane solution is also not a suitable solution. As the worldvolume dimension increases, the spacetime dimension decreases.
The covariant derivative has the following components
\bea
\dd_{ij}&=&P_{ij}
~~,~~
\dd_{\alpha i}~=~0
~~,\dd_{\alpha\beta}~=~\frac{1}{2}\epsilon_{\alpha  \beta\gamma}\epsilon_{jk}\partial^\gamma X^{jk}~~~.
\eea
The Virasoro operator has the following non-trivial component
\bea
{\cal S}^\alpha&=&\frac{1}{8}\dd_{\beta \gamma}\epsilon^{\alpha\beta\gamma}\epsilon^{ij}\dd_{ij}~=~
\frac{1}{4}(\partial^\alpha X^{ij})P_{ij}~~,~~
{\cal S}^{i}~=~0 ~~~,
\eea
which satisfy the closed algebra
\bea
\left[{\cal S}^\alpha(\sigma), {\cal S}^\beta(\sigma') \right]&=&i\left\{
{\cal S}^{(\alpha}(\sigma) \partial^{\beta)}\delta(\sigma-\sigma')
+{\cal S}^{(\alpha}(\sigma') \partial^{\beta)}\delta(\sigma-\sigma')
\right\}\nn\\
	&&-i\partial^{[\alpha}{\cal S}^{\beta]}\delta(\sigma-\sigma')
~~~.
\eea
The Virasoro constraint ${\cal S}^\alpha=0$ generates the  reparametrization of the worldvolume $\sigma_\alpha$ coordinate
as in the membrane solution.
This is an inconsistency of the 3-brane solution of the Gauss law constraint.
The southwest  sectioning of the ${\cal A}$-theory diamond \bref{AGATMSWeb}  can not be performed, since 
 both the Virasoro dimensional reduction condition 
and the section condition have no
nontrivial components in the same way as in the membrane solution.
Therefore it is not suitable as a duality manifest brane theory.

\end{enumerate}

\vskip 6mm
\subsection{D=4: SO(5,5) case}

The structure of the Gauss law constraint, ${\cal U}_\mu= P\slashed{\partial}=0$, and  ${\cal V}=\partial^2=0$ with
$\slashed{\partial}=\partial^m \gamma_m$ is similar to that of the $\kappa$-symmetry constraint of the superparticle, $d\slashed{p}=0=p^2$ with
$\slashed{p}=p^m \gamma_m$.  
Therefore the strategy is to first solve ${\cal V}=0$ in the lightcone gauge, then solve ${\cal U}=0$ as the lightcone projected spinor. 
The spacetime coordinate is the Majorana-Weyl spinor of the SO(5,5).
The spinor representation of SO(5,5) is the same as SO(9,1)
since spinor representations of SO(D$_+$,D$_-$) depend on  D$_++$D$_-$ mod 8 and D$_-$ mod 4 ( for example, see Ref. \cite{Siegel:1988yz}).
For the D=4 SO(5,5) case, a brane theory obtained from the 10-brane in the 16-dimensional spacetime by solving the Gauss law dimensional reduction condition must have the spacetime dimension larger than four.
We demonstrate that enlarging the worldvolume leads to a stronger restriction on the spacetime imposed by the Gauss law dimensional reduction condition. Each enlargement of the worldvolume halves the spacetime.

\begin{enumerate}
\item{{\bf String}
		
At first we solve ${\cal V}=0$ in the lightcone gauge 
\cite{Linch:2015qva},
where the string worldsheet is taken in the lightcone direction as $\sigma_+=\sigma$, $\sigma_\pm~=~\frac{1}{2}(\sigma_1\pm \sigma_6)$ as
\bea
{\cal V}&=&\hat{\eta}_{mn}\partial^m\partial^n
~=~2\partial^+\partial^-+\partial^i\partial^i~=~0~~~\nn\\
\partial^+&=&\partial_\sigma~\neq~0~~,~~\partial^-=\partial^i~=~0~~,~~i=(2,\cdots,5;7\cdots,10)\label{lcg}~~~\\
\partial^\pm&=&\frac{1}{2}(\partial^1\pm\partial^6)~~,~~
\hat{\eta}^{mn}=(1,\cdots,1;-1\cdots,-1)\nn~~~.
\eea
Then next we solve the Gauss law dimensional reduction condition $\overline{\cal U}_\mu=0$.
The gamma matrices are given in the lightcone indices as
\bea
\gamma^\pm&=& \frac{1}{\sqrt{2}}(\gamma^1\pm  \gamma^6)=\frac{1}{\sqrt{2}}\gamma^1(1\pm \gamma^1\gamma^6)~~~\nn\\
{\cal P}^{\pm}&=&
\frac{1}{2}\gamma^\pm\gamma^\mp=\frac{1}{2}(1\mp\gamma^1\gamma^6)~~~ ~~~
\eea
with ${\cal P}^++{\cal P}^-=1$,
 ${\cal P}^\pm{\cal P}^{\pm}={\cal P}^{\pm}$, 
 ${\cal P}^+{\cal P}^-=0={\cal P}^-{\cal P}^+$
and the Gauss law dimensional reduction condition is solved in the lightcone gauge \bref{lcg}  as
\bea
\overline{\cal U}_\mu&=&
P(\partial^+\gamma^- +\partial^-\gamma^+ + \partial^i\gamma^i )~=~0\nn\\
\Rightarrow&&P\gamma^-~=~0~~\to~~P~=~P{\cal P}^+~\equiv~P^+\label{lcP}~~~.
\eea
The spacetime coordinate is also consistently projected as
\bea
X&=&{\cal P}^+X~\equiv~X^+~~,~~\left[P^+{}_\mu(\sigma),X^+{}^\nu(\sigma')\right]=\frac{1}{i}{\cal P}^+{}^\nu{}_\mu\delta(\sigma-\sigma')~~~.
\eea
The spacetime is reduced to 8 dimensions. This solution is
a ${\cal T}$-string in the O(4,4) spacetime.
The covariant derivative becomes 
\bea
\dd_\mu&=& P^++
\overline{\partial^+{X}^+}\gamma^-~=~
\dd_\nu{\cal P}^+{}^\nu{}_\mu~\equiv~\dd^+{}_\mu~~,~~\dd^-{}_\mu~=~0\label{covderi}
\eea
where $({\cal P}^\pm)^T=C~{\cal P}^\mp~ C^{-1}$
 is used.
The Virasoro operator has the following non-trivial component
\bea
{\cal S}^+&=&\frac{1}{4}\bar{\dd}^+\gamma^+\dd^+~~,~~
{\cal S}^-~=~
{\cal S}^i~=~0\label{TstringVirasoro}~~~.
\eea
The Virasoro constraint ${\cal S}^+=0$ generates the diffeomorphism in $\sigma$ direction. In the Virasoro dimensional reduction condition one of  $P^+{}_\mu(\sigma)$s is  replaced with its 0-mode $p^+{}_\mu$
\bea
\overline{\cal S}^+&=&\frac{1}{2} \bar{p}^+ \gamma^+ P^+(\sigma)=0~~~.
\eea
The section condition is the Virasoro constraint in which both  $P^+(\sigma)$s are replaced  with 0-modes 
\bea
\underline{\cal S}^+&=&\frac{1}{2} \bar{p}^+\gamma^+{} p^+=0
\eea
This string solution \bref{lcg} is a consistent solution of the Gauss law dimensional reduction condition representing the ${\cal T}$-string in O(4,4) spacetime. The diffeomorphism of the string worldsheet is generated by  \bref{TstringVirasoro}. The string in the 4-dimensional spacetime is obtained by further sectioning \cite{Hatsuda:2024kuw}
.}

\item{ {\bf Membrane}

 {Since the 10-dimensional N=(1,0) supersymmetry algebra contains no membrane two form charge, it does not admit a membrane type half BPS projection on the 16-component Majorana–Weyl spinor.}
We solve the constraint ${\cal V}=0$ by identifying the doubled lightcone directions,  $\sigma_+$ and $\sigma_{+'}$, as the membrane worldvolume coordinates with $\sigma_{\pm'}~=~\frac{1}{2}(\sigma_2\pm \sigma_7)$.
\bea
{\cal V}&=&2\partial^+\partial^-+2\partial^{+'}\partial^{-'}+
\partial^i\partial^i=0\nn\\
\partial^+&\neq &
0~~,~~\partial^{+'}
~\neq~0~~,~~\partial^-=\partial^{-'}=\partial^i~=~0~~\label{dlcg}~~~\\
\partial^{\pm '}&=&\frac{1}{2}(\partial^2\pm\partial^7)~~,~~i=(3,4,5; 8,9,10)~~~.\nn
\eea
The gamma matrices are given as
\bea
\gamma^{\pm '}&=& \frac{1}{\sqrt{2}}(\gamma^2\pm \gamma^7)=\frac{1}{\sqrt{2}}\gamma^2(1\pm \gamma^2\gamma^7)~~~\nn\\
{\cal P}^{\pm '}&=&\frac{1}{2}\gamma^{\pm'}\gamma^{\mp'}=\frac{1}{2}(1\mp\gamma^2\gamma^7)~~~
\eea
which satisfy 
${\cal P}^{+'}+{\cal P}^{-'}
=1$,  ${\cal P}^{\pm'}{\cal P}^{\pm '}={\cal P}^{\pm '}$, ${\cal P}^{+'}{\cal P}^{-'}=0={\cal P}^{-'}{\cal P}^{+'}$,
$\left[{\cal P}^\pm,{\cal P}^{\pm '}
\right]=0$.
The Gauss law {dimensional reduction} condition is solved as
\bea
\overline{\cal U}_\mu&=&P
\left(\gamma^+\partial^-
+\gamma^-\partial^+ 
+\gamma^{+'}\partial^{-'}
+\gamma^{-'}\partial^{+'}
+\gamma^i\partial^i
\right)~=~0\nn\\
&&~~\Rightarrow~P\gamma^- =P\gamma^{-'} =0~~,~~P=
P{\cal P}^+{\cal P}^{+'}=P^{++'}~~~.
\eea
The spacetime coordinate is $X={\cal P}^+{\cal P}^{+'}X\equiv X^{++'}$, and the spacetime dimension is 4.  
Since the spacetime dimension must be larger than 4, the membrane solution is not a suitable solution.
The covariant derivative becomes
\bea
\dd_\mu&=&\dd^{++'}{}_\mu~=~P^{++'}~~,~~\dd^{+-'}{}_\mu~=~\dd^{-+'}{}_\mu~=~\dd^{--'}{}_\mu~=~0
\eea
since the doubled projection operator eliminates the $\partial X$ term by $\overline {\partial^+X}{\cal P}^-{\cal P}^{-'}\gamma^-
{\cal P}^+{\cal P}^{+'}=0$ and  
$\overline {\partial^{+'}X}{\cal P}^-{\cal P}^{-'}\gamma^{-'}
{\cal P}^+{\cal P}^{+'}
=0$.
The Virasoro operator has no non-trivial component
\bea
{\cal S}^m&=&0
\label{MembraneVirasoro}~~~
\eea
from ${\cal P}^-
{\cal P}^{-'}\gamma^{m}{\cal P}^+{\cal P}^{+'}=0$.
This membrane solution does not possess the worldvolume diffeomorphism, so the string worldsheet diffeomorphism is not included. 
Therefore it is not suitable as a duality manifest brane theory.}
\end{enumerate}
\par
\vskip 6mm

\subsection{D$\geq$5: E$_{\rm D+1}$ case}

The D=5 case and the D=6 case were examined in \cite{Siegel:2018puf} and \cite{Siegel:2020qef}. 
In both cases the  additional constraint $\mathcal{V}=0$ determines the solution.
For the D=5 case the string solution is a consistent solution and other solutions are unlikely to be  allowed, while
for the D=6 case further analysis is required.

\vskip 6mm
\section{Covariantized string solution and conformal symmetry}\label{section:5}
\subsection{Covariantized string solution of the Gauss law constraint}
We have shown that the only consistent solution of the Gauss law dimensional reduction condition is the string worldsheet for the D=3 and D=4 cases.
The string solution of the Gauss law dimensional reduction condition 
$\overline{\cal U}_{\cal M}={\mathbbm h}_{1{\cal M}}^{M}\partial^1 P_{M}=0$ is also a string solution of the Gauss law constraint
$\overline{\cal U}_{\cal M}=0$ at least for D=3, 4.
A constant exceptional group rotation 
creates the exceptional group  covariant vector $q^m$
as
\bea
\partial^5=\partial_\sigma ~\to~\partial^5
\Lambda_5{}^m~=~q^m\partial_\sigma~~~.
\eea
We propose a covariantized solution of the Gauss law constraint as
\bea
\partial^m&=&
q^m\partial_\sigma~\label{derivative1string} 
\eea
where $q^m$ is a normalized constant vector satisfying $~q^mq_m=1~$.
The spacetime is also transformed  under the exceptional symmetry transformation, while still satisfying the Gauss law constraint,
\bea
P_M &\to&  P'_M~=~{\cal R}(\Lambda)P_M \nn\\
{\rm with}~~
{\cal U}_{\cal M}|_{1}=
{\mathbbm h}^{M}_{1{\cal M}}\partial^1 P_M~=0
&\to&
{\cal U}_{\cal M}|_{q}=
{\mathbbm h}^{M}_{m{\cal M}}q^m\partial_\sigma P'_M=0~~~.\label{qU}
\eea

{For the D=3 case, both the worldvolume coordinate and the spacetime coordinate are covariantized under SL(5) as follows.
The string solution is given by \bref{stringD3} and \bref{ststringD3}
\bea
&&\partial^5~=~\partial_\sigma\neq 0~,~\partial^i~=~0~~,i=1,\cdots,4\nn\\
&&P_{5i}~=~0~~,~~P~=~P_{ij}~=~-P_{ji}~~~,
\eea
while the covariantized solution is given as
\bea
\to~~
&&\partial'^m=(\partial'^5,\partial'^i)~~,~~\partial'^5~=~q^5\partial_\sigma\neq 0~,~\partial'^i~=~q^i\partial_\sigma~~,~~q^5\neq 0\nn\\
&&P'_{mn}=(P'_{5i},P'_{ij})~~,~~P'_{5i}~=~-\displaystyle\frac{1}{q^5}q^j \lambda_{j}{}^k   \lambda_i{}^l  P_{kl}~~,~~
P'_{ij}~=~\lambda_i{}^k \lambda_j{}^l P_{kl} 
\\
&&{\rm SL}(5)~\ni~\Lambda_m{}^n=\left(
\begin{array}{cc}
(\lambda^{-1})_i{}^j&0\\q^j&q^5\end{array}
\right)~~,~~{\rm det}~\lambda_{i}{}^j={q^5}~~~.\nn
\eea
In this way, both the brane coordinate and  the spacetime coordinate are  covariantized under the exceptional group.}

\vskip 6mm

\subsection{Relation to the constant charge parameter in the Exceptional $\sigma$-model}

Arvanitakis and Blair showed that the Lagrangian of the two-dimensional $\sigma$-model including the Wess-Zumino term 
with    $q_{MN}$ as the constant charge parameter  \cite{Arvanitakis:2018hfn}.  $q_{MN}$ satisfies the following constraint given in  eq. (1.4) of \cite{Arvanitakis:2018hfn} 
\bea
q_{MN}Y^{NK}_{PQ}\dd_K=q_{PQ}\dd_M~~~.\label{ESM}
\eea
The  {exceptional group invariant tensor}  $Y^{NK}_{PQ}$ \cite{Berman:2012uy}
is related to the exceptional group invariant metric $\eta^{MNm}$  as \bref{braneee}
\bea
Y^{NK}_{PQ}=\eta^{NKm}\eta_{PQm}~~~.
\eea
Let us relate the constant charge parameter $q_{MN}$ to the constant worldvolume vector $q^m$ of the covariantized string solution \bref{derivative1string} as
\bea
q_{MN}~=~\eta_{MNm}q^m~~~.
\eea
The left hand side of  \bref{ESM} becomes
\bea
\eta_{MNn}q^n \eta^{NKm}\eta_{PQm}\dd_K&=&
\eta_{PQn}q^n\dd_M-\eta_{PQn}U_{Mm}^{Kn}q^m\dd_K\nn\\
&=&\eta_{PQn}q^n\dd_M
-{\mathbbm h}_M^{n{\cal M}}\eta_{PQn} ({\mathbbm h}_{m{\cal M}}^K q^m\dd_K)~~~.
\eea
The right hand side of \bref{ESM} is
$q\eta_{PQn}^n\dd_M$, then the constraint \bref{ESM} reduces to 
\bea
\Rightarrow~~ {\mathbbm h}_{m{\cal M}}^K q^m\dd_K&=&0~~~.
\eea
This is a coefficient of the worldvolume derivative in the  Gauss law constraint with the covariantized string solution \bref{derivative1string}.
The constant charge parameter $q_{MN}=\eta_{MNm}q^m$ in the Exceptional $\sigma$-model corresponds to the worldvolume vector $q^m$ in the covariantized string solution of the Gauss law constraint.
This relation suggests that  
the constant charge parameter effectively encodes
the worldvolume structure governed by the Gauss law constraint.
\par
\vskip 6mm
\subsection{Conformal symmetry}

It is easy to show that the brane Virasoro algebra includes the standard Virasoro algebra by using the covariantized string solution \bref{derivative1string}, thereby naturally embedding the two-dimensional conformal symmetry.
Let us contract  the brane Virasoro constraint   \bref{BraneVirasoro}  with the $q_m$ in \bref{qU}
to define $\hat{\cal S}$ 
\bea
{\cal S}^mq_m=\hat{\cal S}~~~.
\eea
The commutator of $\hat{\cal S}$'s is obtained by multiplying  $q_mq_n$ on
the Virasoro algebra  \bref{SSSU} 
and using \bref{qU}  {together with commutators with ${\cal H}$'s} as 
\bea
\left[ \hat{\cal S}(\sigma),\hat{\cal S}(\sigma')\right]
&=&
i\left(\hat{\cal S}(\sigma)+\hat{\cal S}(\sigma')\right)\partial_\sigma\delta(\sigma-\sigma')\nn
\\
 {\left[ \hat{\cal S}(\sigma),{\cal H}(\sigma')\right]}
&=& {{i}\left({\cal H}(\sigma)+{\cal H}(\sigma')\right)\partial_\sigma\delta(\sigma-\sigma')}~\label{2Dconformal}\\
 {\left[ {\cal H}(\sigma),{\cal H}(\sigma')\right]}
&=& {{i}\left(\hat{\cal S}(\sigma)+\hat{\cal S}(\sigma')\right)\partial_\sigma\delta(\sigma-\sigma')}~~~.\nn
\eea
This is  the standard Virasoro algebra generating the conformal symmetry of the string worldsheet \bref{stringVirasoro}.
The conformal symmetries of ${\cal A}$-theory branes 
exhibit precisely the two-dimensional conformal symmetry after solving the Gauss law constraint.

\par\vskip 6mm
\section{Discussion}\label{section:6}

The Gauss law constraint in ${\cal A}$-theory branes promotes spacetime coordinates to gauge fields and enlarges the string worldsheet into the brane worldvolume.
We have shown that the consistent solution of the
Gauss law constraint is the only string solution for D=3 and 4 cases,  {as a free object with quantizable bilinear form Lagrangians}. 
 {Branes with nonlinear form Lagrangians such as M2-brane are not considered here.}
We also proposed the covariantized string solution $\partial^m=q^m \partial_\sigma$  in such a way that it solves the Gauss law constraint,  ${\cal U}_{\cal M}={\mathbbm h}_{m{\cal M}}^M\partial^m\dd_M={\mathbbm h}_{m{\cal M}}^M q^m \partial_\sigma\dd_M =0$.
In this gauge the brane Virasoro algebra reduces to the standard Virasoro algebra, thereby exhibiting worldsheet conformal symmetry.
Once the physical worldvolume is chosen, the quantization is performed analogously to the standard string theory.
We leave the quantization of ${\cal A}$-theory branes for future work.

The two-dimensional conformal symmetry implies that the scattering amplitudes are described by the dual resonance model.
Since the Gauss law constraint reduces massive brane modes to the two-dimensional massive string modes,  { $X(\tau,\sigma_m)$ $\to$ $X(\tau,\sigma)$}, while leaving massless modes intact, one could have an educated guess of the structure of ${\cal{A}}$-theory.
Schematically ${\cal{A}}$-theory amplitude can be written as $A_{n} = \langle\, V_{1}, \cdots V_{n}\, \rangle$. $V$ is the vertex operators, which can be written as $V = E(\epsilon, \partial X) e^{i k \cdot X}$. $E(\epsilon, \partial X)$ is a function of the polarization vector $\epsilon$ and $\partial X$.

When one evaluates $A_{n}$,  {contraction between $\partial X$'s and $e^{ik\cdot X}$ operators and contraction among $\partial X$'s themselves give the kinematic factor of the amplitudes. While the contraction between $e^{ik_{i}\cdot X}$ and $e^{ik_{j}\cdot X}$ gives the dynamic factors (where the massive poles are). Since the $XX$ correlators coincide with those of string theory after solving the Gauss law constraint, we expect the dynamical factors to be  identical to those in string theory. }

As an example, for D=3 case, 4-point ${\cal{A}}$-theory amplitudes is written as  \cite{Siegel:2020gro}
\[
   A_{4}= K(\epsilon_{1}, \epsilon_{2}, \epsilon_{3}, \epsilon_{4}) \frac{\Gamma(-s)\Gamma(-t)\Gamma(-u)}{\Gamma(1+ s)\Gamma(1+ t)\Gamma(1+ u)},
\]
where $K(\epsilon_{1},\cdots\epsilon_{4})$ is the kinematic factor which is the function of polarization vectors and momentum $k_{i}$. By using its subgroup  $H$=Sp(4)  covariance, gauge invariance plus the section condition, one could completely fix the kinematic factor $K$, which is given in \cite{Siegel:2020gro}. We hope to more systematically bootstrap the amplitude using this method in the future.

Since the  $\partial^m$ is the canonical conjugate of $\sigma_m$, the Gauss law  constraint generates the reparametrization of the worldvolume coordinate   as 
\bea
\delta_\lambda \sigma_m&=&
\left[\frac{1}{i}\displaystyle\int \lambda^{\cal M}\vec{\cal U}_{\cal M}, \sigma_m\right]~=~
\frac{1}{i}\lambda^{\cal M}{\mathbbm h}^M_{m{\cal M}}\dd_M~~~.
\eea
The vector notation  $\vec{\cal U}_{\cal M}$ denotes
the differential operator acting on an arbitrary function as in 
\bref{Gausslaw}, \bref{sectionU}, \bref{Gausssection}.
Under the gauge transformation generated 
by the Gauss law constraint, 
the worldvolume coordinate is also shifted for the D=3 case as
\bea
\delta_\lambda \sigma_m&=&
\frac{1}{i}\lambda^{n}\dd_{mn}~~~,
\eea
and for the D=4 case as
\bea
\delta_\lambda \sigma_m&=&
\frac{1}{i}\lambda^{\nu}\gamma_m{}^\mu{}_{\nu}\dd_\mu
~~~.
\eea
To eliminate these gauge degrees of freedom covariantly,
the gauge fixing is required.

The covariantized solution of the Gauss law constraint in \bref{derivative1string} 
corresponds to the string worldsheet gauge fixing $\sigma_5\neq 0$ and $\sigma_i=0~~\rm{for} ~i\neq 5$ 
 {with $\hat{m}=0,1,\cdots,5$}
as
\bea
\sigma_{\hat{m}}-\sigma_5\delta_{{\hat{m}}5}&=&0~~~.
\eea
To use this condition in the path integral measure,
this is written in a covariant way with $\sigma_5\to (\sigma_{\parallel}){}_{\hat{m}}$ {$=(0,\cdots,\sigma_5)$} as
\bea
\sigma_{\hat{m}}=(\sigma {\cal P}_{\parallel})_{\hat{m}}~~{\rm with}~~~
{\cal P}_\parallel{}^{\hat{n}}{}_{\hat{m}}=\delta^{\hat{n}}{}_{\hat{m}}-
\frac{\sigma_{\parallel}{}^{\hat{n}} \sigma_{\parallel}{}_{\hat{m}}
}{\sigma_{\parallel}{}^{\hat{l}} \sigma_{\parallel}{}_{\hat{l}}}~~~.
\eea
The gauge fixing  condition and the gauge generator are imposed in the worldvolume measure as
\bea
\int d^6\sigma_{\hat{m}} d^6k^{\hat{m}} \delta(\dd_{nl}k^l)[{\rm Det}\dd_{mn}]\delta(\sigma_{\hat{n}}-\sigma_5\delta_{{\hat{n}}5}){\cal L}(\sigma)~~~
\eea
where the determinant factor comes from the Jacobian of the gauge symmetry transformation of the gauge fixing condition
\bea
\frac{\delta \delta_\zeta\left( \sigma_{\hat{m}}-\sigma_5\delta_{{\hat{m}}5}\right)}{\delta \zeta^{\hat{n}} }=\dd_{\hat{m}\hat{n}}~~~.
\eea

However this gauge choice breaks the gauge symmetry of the current, and as a result the exceptional symmetry covariant selfduality is obscured.
To quantize the theory with manifest exceptional  duality symmetry, 
we will likely need a BRST quantization that incorporates both the first-quantized ghosts and the “zeroth-quantized” ghost
 \cite{Siegel:2016dek}.
We leave the  covariant quantization of the ${\cal A}$-theory brane  for future work as well.

In this paper, non-linear solutions of the Gauss law constraint are not examined. In our previous work \cite{Hatsuda:2024kuw}, the standard M2-brane solution of supergravity was derived from the ${\cal A}$-theory brane by the Virasoro sectioning together with a non-linear projection.
The derivation of other standard branes, including D-branes, may involve a non-linear form of the Gauss law constraint.
A concrete derivation is left for future work.

\par\vskip 6mm

\subsection*{Acknowledgments}

We would like to express our deepest gratitude to Warren Siegel for his profound insights and invaluable discussions that have shaped this project at every stage.
We also appreciate his unpublished draft “Furniture,” which has influenced the direction of this research.
 {We are grateful to Yuho Sakatani  for many valuable and helpful comments.}
We also acknowledge the Simons Center for Geometry and Physics for its hospitality during
``The Simons Summer Workshop in Mathematics and Physics 2025" 
where this work was developed.
M.H. is supported in part by 
Grant-in-Aid for Scientific Research (C), JSPS KAKENHI
Grant Numbers JP22K03603, JP25K07324, JP25K06999.
\printbibliography

@article{Duff:2015jka,
    author = "Duff, M. J. and Lu, J. X. and Percacci, R. and Pope, C. N. and Samtleben, H. and Sezgin, E.",
    title = "{Membrane Duality Revisited}",
    eprint = "1509.02915",
    archivePrefix = "arXiv",
    primaryClass = "hep-th",
    reportNumber = "USTC-ICTS-15-08, MI-TH-1529",
    doi = "10.1016/j.nuclphysb.2015.10.003",
    journal = "Nucl. Phys. B",
    volume = "901",
    pages = "1--21",
    year = "2015"
}

@article{Berman:2012uy,
    author = "Berman, David S. and Musaev, Edvard T. and Thompson, Daniel C. and Thompson, Daniel C.",
    title = "{Duality Invariant M-theory: Gauged supergravities and Scherk-Schwarz reductions}",
    eprint = "1208.0020",
    archivePrefix = "arXiv",
    primaryClass = "hep-th",
    doi = "10.1007/JHEP10(2012)174",
    journal = "JHEP",
    volume = "10",
    pages = "174",
    year = "2012"
}

@article{Obers:1998fb,
    author = "Obers, N. A. and Pioline, B.",
    title = "{U duality and M theory}",
    eprint = "hep-th/9809039",
    archivePrefix = "arXiv",
    reportNumber = "CERN-TH-98-282, CPHT-S639-0898",
    doi = "10.1016/S0370-1573(99)00004-6",
    journal = "Phys. Rept.",
    volume = "318",
    pages = "113--225",
    year = "1999"
}

@article{Siegel:2020gro,
    author = "Siegel, Warren and Wang, Yu-Ping",
    title = "{F-theory amplitudes}",
    eprint = "2010.14590",
    archivePrefix = "arXiv",
    primaryClass = "hep-th",
    month = "10",
    year = "2020"
}

@book{Siegel:1988yz,
  author = "Siegel, Warren",
  title = "{Introduction to string field theory}",
  eprint = "hep-th/0107094",
  archivePrefix = "arXiv",
  reportNumber = "YITP-SB-01-39",
  volume = "8",
  year = "1988",
}

@article{Hatsuda:2024kuw,
  author = "Hatsuda, Machiko and Hul{\'\i}k, Ond{\v{r}}ej and Linch, William D.
            and Siegel, Warren D. and Wang, Di and Wang, Yu-Ping",
  title = "{Strings and membranes from $\mathcal{A}$-theory five brane}",
  eprint = "2410.11197",
  archivePrefix = "arXiv",
  primaryClass = "hep-th",
  doi = "10.21468/SciPostPhys.19.1.009",
  journal = "SciPost Phys.",
  volume = "19",
  number = "1",
  pages = "009",
  year = "2025",
}

@article{Berman:2011cg,
  author = "Berman, David S. and Godazgar, Hadi and Godazgar, Mahdi and Perry,
            Malcolm J.",
  title = "{The Local symmetries of M-theory and their formulation in
           generalised geometry}",
  eprint = "1110.3930",
  archivePrefix = "arXiv",
  primaryClass = "hep-th",
  reportNumber = "QMUL-PH-11-15, DAMTP-2011-86",
  doi = "10.1007/JHEP01(2012)012",
  journal = "JHEP",
  volume = "01",
  pages = "012",
  year = "2012",
}

@article{Coimbra:2011ky,
  author = "Coimbra, Andr{\'e} and Strickland-Constable, Charles and Waldram,
            Daniel",
  title = "{$E_{d(d)} \times \mathbb{R}^+$ generalised geometry, connections and
           M theory}",
  eprint = "1112.3989",
  archivePrefix = "arXiv",
  primaryClass = "hep-th",
  reportNumber = "IMPERIAL-TP-11-DW-02",
  doi = "10.1007/JHEP02(2014)054",
  journal = "JHEP",
  volume = "02",
  pages = "054",
  year = "2014",
}

@article{Hohm:2010pp,
  author = "Hohm, Olaf and Hull, Chris and Zwiebach, Barton",
  title = "{Generalized metric formulation of double field theory}",
  eprint = "1006.4823",
  archivePrefix = "arXiv",
  primaryClass = "hep-th",
  reportNumber = "IMPERIAL-TP-2010-CH-03, MIT-CTP-4154",
  doi = "10.1007/JHEP08(2010)008",
  journal = "JHEP",
  volume = "08",
  pages = "008",
  year = "2010",
}

@article{Kugo:1992md,
  author = "Kugo, Taichiro and Zwiebach, Barton",
  title = "{Target space duality as a symmetry of string field theory}",
  eprint = "hep-th/9201040",
  archivePrefix = "arXiv",
  reportNumber = "YITP-K-961, IASSNS-HEP-92-3, MIT-CTP-2058",
  doi = "10.1143/ptp/87.4.801",
  journal = "Prog. Theor. Phys.",
  volume = "87",
  pages = "801--860",
  year = "1992",
}

@article{Yamatsu:2015npn,
  author = "Yamatsu, Naoki",
  title = "{Finite-Dimensional Lie Algebras and Their Representations for
           Unified Model Building}",
  eprint = "1511.08771",
  archivePrefix = "arXiv",
  primaryClass = "hep-ph",
  reportNumber = "OU-HET 886, KYUSHU-HET-216, OU-HET-886",
  month = "11",
  year = "2015",
}

@article{Siegel:2016dek,
  author = "Siegel, W.",
  title = "{F-theory with zeroth-quantized ghosts}",
  eprint = "1601.03953",
  archivePrefix = "arXiv",
  primaryClass = "hep-th",
  reportNumber = "YITP-SB-16-2",
  month = "1",
  year = "2016",
}

@article{Blair:2023noj,
  author = "Blair, Chris D. A. and Lahnsteiner, Johannes and Obers, Niels A. and
            Yan, Ziqi",
  title = "{Unification of Decoupling Limits in String and M Theory}",
  eprint = "2311.10564",
  archivePrefix = "arXiv",
  primaryClass = "hep-th",
  reportNumber = "NORDITA-2023-071, IFT-UAM/CSIC-23-151",
  doi = "10.1103/PhysRevLett.132.161603",
  journal = "Phys. Rev. Lett.",
  volume = "132",
  number = "16",
  pages = "161603",
  year = "2024",
}

@article{Blair:2019tww,
  author = "Blair, Chris D. A.",
  title = "{Open exceptional strings and D-branes}",
  eprint = "1904.06714",
  archivePrefix = "arXiv",
  primaryClass = "hep-th",
  doi = "10.1007/JHEP07(2019)083",
  journal = "JHEP",
  volume = "07",
  pages = "083",
  year = "2019",
}

@article{Arvanitakis:2017hwb,
  author = "Arvanitakis, Alex S. and Blair, Chris D. A.",
  title = "{Unifying Type-II Strings by Exceptional Groups}",
  eprint = "1712.07115",
  archivePrefix = "arXiv",
  primaryClass = "hep-th",
  reportNumber = "IMPERIAL-TP-2017-ASA-02",
  doi = "10.1103/PhysRevLett.120.211601",
  journal = "Phys. Rev. Lett.",
  volume = "120",
  number = "21",
  pages = "211601",
  year = "2018",
}

@article{Sakatani:2016sko,
    author = "Sakatani, Yuho and Uehara, Shozo",
    title = "{Branes in Extended Spacetime: Brane Worldvolume Theory Based on Duality Symmetry}",
    eprint = "1607.04265",
    archivePrefix = "arXiv",
    primaryClass = "hep-th",
    doi = "10.1103/PhysRevLett.117.191601",
    journal = "Phys. Rev. Lett.",
    volume = "117",
    number = "19",
    pages = "191601",
    year = "2016"
}

@article{Sakatani:2017xcn,
    author = "Sakatani, Yuho and Uehara, Shozo",
    title = "{{\ensuremath{\eta}}-symbols in exceptional field theory}",
    eprint = "1708.06342",
    archivePrefix = "arXiv",
    primaryClass = "hep-th",
    doi = "10.1093/ptep/ptx151",
    journal = "PTEP",
    volume = "2017",
    number = "11",
    pages = "113B01",
    year = "2017"
}

@article{Sakatani:2017vbd,
    author = "Sakatani, Yuho and Uehara, Shozo",
    title = "{Exceptional M-brane sigma models and $\eta$-symbols}",
    eprint = "1712.10316",
    archivePrefix = "arXiv",
    primaryClass = "hep-th",
    doi = "10.1093/ptep/pty021",
    journal = "PTEP",
    volume = "2018",
    number = "3",
    pages = "033B05",
    year = "2018"
}

@article{Sakatani:2022auu,
  author = "Sakatani, Yuho and Uehara, Shozo",
  title = "{Gauged sigma models and exceptional dressing cosets}",
  eprint = "2203.16532",
  archivePrefix = "arXiv",
  primaryClass = "hep-th",
  doi = "10.1093/ptep/ptac098",
  journal = "PTEP",
  volume = "2022",
  number = "9",
  pages = "093B01",
  year = "2022",
}

@article{Osten:2024mjt,
  author = "Osten, David",
  title = "{On the universal exceptional structure of world-volume theories in
           string and M-theory}",
  eprint = "2402.10269",
  archivePrefix = "arXiv",
  primaryClass = "hep-th",
  doi = "10.1016/j.physletb.2024.138814",
  journal = "Phys. Lett. B",
  volume = "855",
  pages = "138814",
  year = "2024",
}

@article{Arvanitakis:2018hfn,
  author = "Arvanitakis, Alex S. and Blair, Chris D. A.",
  title = "{The Exceptional Sigma Model}",
  eprint = "1802.00442",
  archivePrefix = "arXiv",
  primaryClass = "hep-th",
  reportNumber = "IMPERIAL-TP-2018-ASA-01",
  doi = "10.1007/JHEP04(2018)064",
  journal = "JHEP",
  volume = "04",
  pages = "064",
  year = "2018",
}

@inproceedings{Polchinski:1996na,
  author = "Polchinski, Joseph",
  title = "{Tasi lectures on D-branes}",
  booktitle = "{Theoretical Advanced Study Institute in Elementary Particle
               Physics (TASI 96): Fields, Strings, and Duality}",
  eprint = "hep-th/9611050",
  archivePrefix = "arXiv",
  reportNumber = "NSF-ITP-96-145",
  pages = "293--356",
  month = "11",
  year = "1996",
}

@article{Siegel:1983es,
  author = "Siegel, W.",
  title = "{Manifest Lorentz Invariance Sometimes Requires Nonlinearity}",
  reportNumber = "UCB-PTH-83/22",
  doi = "10.1016/0550-3213(84)90453-X",
  journal = "Nucl. Phys. B",
  volume = "238",
  pages = "307--316",
  year = "1984",
}

@article{Witten:1995im,
  author = "Witten, Edward",
  title = "{Bound states of strings and p-branes}",
  eprint = "hep-th/9510135",
  archivePrefix = "arXiv",
  reportNumber = "IASSNS-HEP-95-83",
  doi = "10.1016/0550-3213(95)00610-9",
  journal = "Nucl. Phys. B",
  volume = "460",
  pages = "335--350",
  year = "1996",
}

@article{Hatsuda:2012uk,
  author = "Hatsuda, Machiko and Kimura, Tetsuji",
  title = "{Canonical approach to Courant brackets for D-branes}",
  eprint = "1203.5499",
  archivePrefix = "arXiv",
  primaryClass = "hep-th",
  reportNumber = "KEK-TH-1529",
  doi = "10.1007/JHEP06(2012)034",
  journal = "JHEP",
  volume = "06",
  pages = "034",
  year = "2012",
}

@article{Hatsuda:1997pq,
  author = "Hatsuda, Machiko and Kamimura, Kiyoshi",
  title = "{Covariant quantization of the super D string}",
  eprint = "hep-th/9708001",
  archivePrefix = "arXiv",
  reportNumber = "TOHO-FP-9756",
  doi = "10.1016/S0550-3213(98)00171-0",
  journal = "Nucl. Phys. B",
  volume = "520",
  pages = "493--512",
  year = "1998",
}

@article{Hatsuda:1998by,
  author = "Hatsuda, Machiko and Kamimura, Kiyoshi",
  title = "{Wess-Zumino actions for IIA D-branes and their supersymmetries}",
  eprint = "hep-th/9804087",
  archivePrefix = "arXiv",
  reportNumber = "TOHO-FP-9759",
  doi = "10.1016/S0550-3213(98)00547-1",
  journal = "Nucl. Phys. B",
  volume = "535",
  pages = "499--511",
  year = "1998",
}

@article{Kamimura:1997ju,
  author = "Kamimura, Kiyoshi and Hatsuda, Machiko",
  title = "{Canonical formulation of IIB D-branes}",
  eprint = "hep-th/9712068",
  archivePrefix = "arXiv",
  reportNumber = "TOHO-FP-9757",
  doi = "10.1016/S0550-3213(98)00415-5",
  journal = "Nucl. Phys. B",
  volume = "527",
  pages = "381--401",
  year = "1998",
}

@article{Hatsuda:2013dya,
  author = "Hatsuda, Machiko and Kamimura, Kiyoshi",
  title = "{M5 algebra and SO(5,5) duality}",
  eprint = "1305.2258",
  archivePrefix = "arXiv",
  primaryClass = "hep-th",
  reportNumber = "KEK-TH-1629",
  doi = "10.1007/JHEP06(2013)095",
  journal = "JHEP",
  volume = "06",
  pages = "095",
  year = "2013",
}

@article{Hatsuda:2023dwx,
  author = "Hatsuda, Machiko and Hul\'\i{}k, Ond\v{r}ej and Linch, William D.
            and Siegel, Warren D. and Wang, Di and Wang, Yu-Ping",
  title = "{$ \mathcal{A} $-theory \textemdash{} A brane world-volume theory
           with manifest U-duality}",
  eprint = "2307.04934",
  archivePrefix = "arXiv",
  primaryClass = "hep-th",
  doi = "10.1007/JHEP10(2023)087",
  journal = "JHEP",
  volume = "10",
  pages = "087",
  year = "2023",
}

@article{Linch:2015fya,
  author = "Linch, III, William D. and Siegel, Warren",
  title = "{F-theory from Fundamental Five-branes}",
  eprint = "1502.00510",
  archivePrefix = "arXiv",
  primaryClass = "hep-th",
  reportNumber = "UMDEPP-015-002, YITP-SB-15-3",
  doi = "10.1007/JHEP02(2021)047",
  journal = "JHEP",
  volume = "02",
  pages = "047",
  year = "2021",
}

@article{Linch:2015qva,
  author = "Linch, William D. and Siegel, Warren",
  title = "{F-theory with Worldvolume Sectioning}",
  eprint = "1503.00940",
  archivePrefix = "arXiv",
  primaryClass = "hep-th",
  reportNumber = "UMDEPP-015-006, YITP-SB-15-6",
  doi = "10.1007/JHEP04(2021)022",
  journal = "JHEP",
  volume = "04",
  pages = "022",
  year = "2021",
}

@article{Linch:2016ipx,
  author = "Linch, William D. and Siegel, Warren",
  title = "{F-brane Dynamics}",
  eprint = "1610.01620",
  archivePrefix = "arXiv",
  primaryClass = "hep-th",
  reportNumber = "MI-TH-1628, YITP-SB-16-38",
  month = "10",
  year = "2016",
}

@article{Linch:2015fca,
  author = "Linch, William D and Siegel, Warren",
  title = "{Critical Super F-theories}",
  eprint = "1507.01669",
  archivePrefix = "arXiv",
  primaryClass = "hep-th",
  month = "7",
  year = "2015",
}

@article{Siegel:2018puf,
  author = "Siegel, Warren and Wang, Di",
  title = "{Enlarged exceptional symmetries of first-quantized F-theory}",
  eprint = "1806.02423",
  archivePrefix = "arXiv",
  primaryClass = "hep-th",
  month = "6",
  year = "2018",
}

@article{Linch:2015lwa,
  author = "Linch, William D. and Siegel, Warren",
  title = "{F-theory superspace}",
  eprint = "1501.02761",
  archivePrefix = "arXiv",
  primaryClass = "hep-th",
  reportNumber = "UMDEPP-015-001, YITP-SB-15-1",
  doi = "10.1007/JHEP03(2021)059",
  journal = "JHEP",
  volume = "03",
  pages = "059",
  year = "2021",
}

@article{Siegel:2019wrr,
  author = "Siegel, Warren and Wang, Di",
  title = "{F-theory superspace backgrounds}",
  eprint = "1910.01710",
  archivePrefix = "arXiv",
  primaryClass = "hep-th",
  reportNumber = "YITP-SB-19-31",
  month = "10",
  year = "2019",
}

@article{Polacek:2014cva,
  author = "Pol\'a\v{c}ek, Martin and Siegel, Warren",
  editor = "Tecchio, Monica and Levin, Daniel",
  title = "{T-duality off shell in 3D Type II superspace}",
  eprint = "1403.6904",
  archivePrefix = "arXiv",
  primaryClass = "hep-th",
  doi = "10.1007/JHEP06(2014)107",
  journal = "JHEP",
  volume = "06",
  pages = "107",
  year = "2014",
}

@article{Hatsuda:2021wpb,
  author = "Hatsuda, Machiko and Siegel, Warren",
  title = "{Perturbative F-theory 10-brane and M-theory 5-brane}",
  eprint = "2107.10568",
  archivePrefix = "arXiv",
  primaryClass = "hep-th",
  doi = "10.1007/JHEP11(2021)201",
  journal = "JHEP",
  volume = "11",
  pages = "201",
  year = "2021",
}

@article{Siegel:2020qef,
  author = "Siegel, Warren and Wang, Di",
  title = "{M Theory from F Theory}",
  eprint = "2010.09564",
  archivePrefix = "arXiv",
  primaryClass = "hep-th",
  month = "10",
  year = "2020",
}

@article{Linch:2017eru,
  author = "Linch, William and Siegel, Warren",
  title = "{F-brane Superspace: The New World Volume}",
  eprint = "1709.03536",
  archivePrefix = "arXiv",
  primaryClass = "hep-th",
  reportNumber = "YITP-SB-17-32",
  month = "9",
  year = "2017",
}

@article{Hatsuda:2015cia,
  author = "Hatsuda, Machiko and Kamimura, Kiyoshi and Siegel, Warren",
  title = "{Type II chiral affine Lie algebras and string actions in doubled
           space}",
  eprint = "1507.03061",
  archivePrefix = "arXiv",
  primaryClass = "hep-th",
  doi = "10.1007/JHEP09(2015)113",
  journal = "JHEP",
  volume = "09",
  pages = "113",
  year = "2015",
}

@inproceedings{Siegel:1993bj,
  author = "Siegel, W.",
  title = "{Manifest duality in low-energy superstrings}",
  booktitle = "{International Conference on Strings 93}",
  eprint = "hep-th/9308133",
  archivePrefix = "arXiv",
  reportNumber = "ITP-SB-93-50",
  month = "9",
  year = "1993",
}

@article{Siegel:1993th,
  author = "Siegel, W.",
  title = "{Superspace duality in low-energy superstrings}",
  eprint = "hep-th/9305073",
  archivePrefix = "arXiv",
  reportNumber = "ITP-SB-93-28",
  doi = "10.1103/PhysRevD.48.2826",
  journal = "Phys. Rev. D",
  volume = "48",
  pages = "2826--2837",
  year = "1993",
}

@article{Siegel:1993xq,
  author = "Siegel, W.",
  title = "{Two vierbein formalism for string inspired axionic gravity}",
  eprint = "hep-th/9302036",
  archivePrefix = "arXiv",
  reportNumber = "ITP-SB-93-2",
  doi = "10.1103/PhysRevD.47.5453",
  journal = "Phys. Rev. D",
  volume = "47",
  pages = "5453--5459",
  year = "1993",
}

@article{Polacek:2013nla,
  author = "Pol\'a\v{c}ek, Martin and Siegel, Warren",
  title = "{Natural curvature for manifest T-duality}",
  eprint = "1308.6350",
  archivePrefix = "arXiv",
  primaryClass = "hep-th",
  doi = "10.1007/JHEP01(2014)026",
  journal = "JHEP",
  volume = "01",
  pages = "026",
  year = "2014",
}
\end{document}